\documentclass[apj]{emulateapj}

\shorttitle{Star Formation in W40}
\shortauthors{Mallick et al.}

\usepackage{subfigure}
\usepackage{amsmath}
\usepackage{natbib}
\usepackage{float}

\usepackage{graphicx} 
\usepackage{lscape}

\usepackage[bookmarks=true,colorlinks]{hyperref}

\usepackage{color}
\definecolor{darkgreen}{rgb}{0,0.5,0}
\definecolor{darkblue}{rgb}{0,0,0.55}
\hypersetup{colorlinks,linkcolor=blue,filecolor=darkgreen,citecolor=blue}

\newcommand{\msun}{$~M_{\odot}$}

\begin{document}

\title{W40 region in the Gould Belt : An embedded cluster and H~{\sc ii} region 
       at the junction of filaments} 

\author{K. K. Mallick}
\affil{Department of Astronomy and Astrophysics, Tata Institute of Fundamental Research, \\
       Homi Bhabha Road, Colaba, Mumbai 400 005, India}
\email{kshitiz@tifr.res.in}

\author{M. S. N. Kumar}
\affil{Centro de Astrof\'{i}sica da Universidade do Porto, Rua das Estrelas, \\
       4150-762 s/n Porto, Portugal}

\author{D. K. Ojha}
\affil{Department of Astronomy and Astrophysics, Tata Institute of Fundamental Research, \\
       Homi Bhabha Road, Colaba, Mumbai 400 005, India} 

\author{Rafael Bachiller}
\affil{Observatorio Astronómico Nacional (IGN), Alfonso XII 3, 28014, Madrid, Spain}
       
\author{M. R. Samal}
\affil{Aix Marseille Universit\'e, CNRS, LAM (Laboratoire d'Astrophysique de Marseille) \\
       UMR 7326, 13388 Marseille, France} 

\and       
       
\author{L. Pirogov}
\affil{Institute of Applied Physics, Russian Academy of Sciences, \\
       46 Uljanov str., Nizhny Novgorod 603950, Russia}

\begin{abstract}
We present a multiwavelength study of the \object{W40} star-forming region using 
infrared (IR) observations in UKIRT \textit{JHK} bands, \textit{Spitzer} IRAC 
bands, and \textit{Herschel} PACS bands; 2.12 $\mu$m H$_2$ narrow-band imaging; 
and radio continuum observations from GMRT (610 and 1280 MHz), in a field of view 
(FoV) of $\sim$ 34\arcmin $\times$ 40\arcmin\,.  Archival \textit{Spitzer} 
observations in conjunction with near-IR (NIR) observations are used to identify 
1162 Class\,II/III and 40 Class I sources in the FoV. The nearest-neighbour stellar 
surface density analysis shows that majority of these young stellar objects (YSOs) 
constitute the embedded cluster centered on the high-mass source IRS\,1A South. 
Some YSOs, predominantly younger population, are distributed along and trace the
filamentary structures at lower stellar surface density. The cluster radius is 
obtained as 0.44\,pc - matching well with the extent of radio emission - with a 
peak density of 650\,pc$^{-2}$. The \textit{JHK} data is used to map the extinction 
in the region which is subsequently used to compute the cloud mass. It has resulted 
in 126\,M$_{\odot}$ and 71\,M$_{\odot}$ for the central cluster and the northern 
IRS\,5 region, respectively. H$_2$ narrow-band imaging displays significant emission, 
which prominently resembles fluorescent emission arising at the borders of dense 
regions. Radio continuum analysis shows this region as having blister morphology,
with the radio peak coinciding with a protostellar source. Free-free emission 
spectral energy distribution (SED) analysis is used to obtain physical parameters 
of the overall photoionized region and the IRS\,5 sub-region. This multiwavelength 
scenario is suggestive of star formation having resulted from merging of multiple 
filaments to form a hub. Star formation seems to have taken place in two successive 
epochs, with the first epoch traced by the central cluster and the high-mass 
star(s) - followed by a second epoch which is spreading into the filaments as 
uncovered by the Class I sources and even younger protostellar sources along the 
filaments. The IRS\,5 H~{\sc ii} region displays indications of swept-up material 
which has possibly led to the formation of protostars. 
\end{abstract}


\section{Introduction}  
\begin{figure}
\centering
\includegraphics[trim={1.5cm 2.5cm 0cm 4.5cm}, clip, scale=0.47]{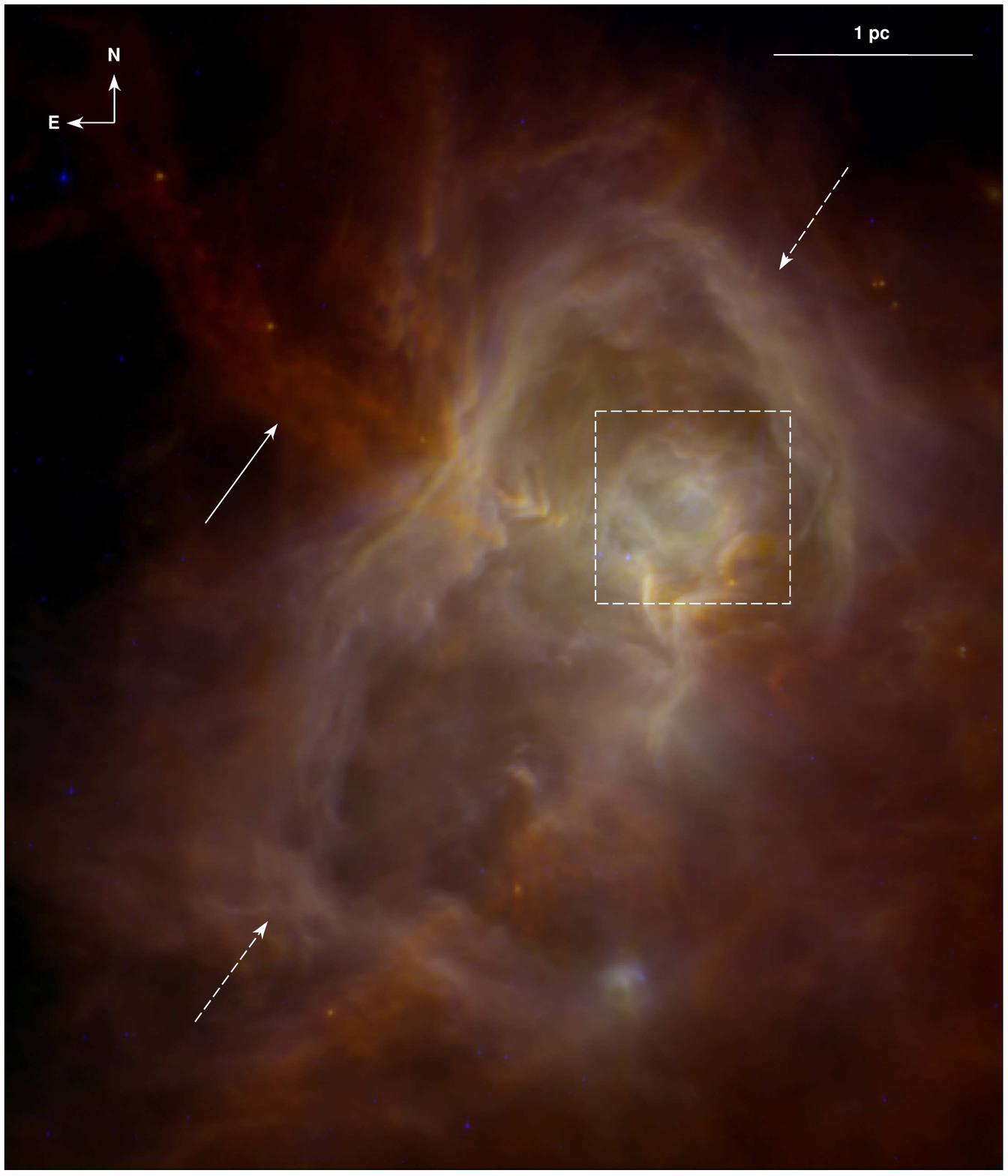}
\caption{Colour composite image of the W40 region 
($\sim$ 34\arcmin $\times$ 40\arcmin) using \textit{Herschel} PACS 160 $\mu$m (red), 
100 $\mu$m (green), and IRAC 8 $\mu$m (blue) images. The dashed white arrows mark 
the two prominent northern and southern lobes of this bipolar nebula. The solid 
white arrow indicates one of the prominent filamentary structures. Dashed white 
box shows the location of a distinct smaller arc-shaped nebula around IRS 5 source.} 
\label{fig_ColourCompositePACS160100IRAC8} 
\end{figure}

\object{W40} \citep{wes58} (also known as \object{Sh2-64}) is an optically 
visible H~{\sc ii} region \citep{fic84} in the Serpens-Aquila Rift region. 
Located at $\alpha_{2000} \sim 18^{h}31^{m}29^{s}$, $\delta_{2000} \sim
-02^{o}05\arcmin\,24\arcsec\,$, it is visible as a bipolar nebula at mid 
and far IR wavelengths; harbouring complex features such as filamentary 
structures and an embedded arc-shaped nebula 
(Figure \ref{fig_ColourCompositePACS160100IRAC8}). 

Early CO observations of this region showed the W40 H~{\sc ii} region to be 
located at the edge of the molecular cloud \object{G28.8+3.5} \citep{zei78}, 
resulting in the well-known blister morphology. Even so, the full extent of 
the molecular gas in this region is not mapped so far. Only the central few 
arcminutes of this cloud are mapped, using isotopologues of CO, finding the 
mass of central cloud core to be $\sim$ 100\msun\, \citep{val92}. W40 also 
has an associated dense molecular clump of $\sim$ 20\arcmin\, diameter 
(\object{TGU 279 P7}; \citealt*[]{dob05}). A large scale, weak molecular 
outflow is also found, through CO observations, to originate in the molecular 
cloud \citep{zhu06}. Located at a Galactic latitude of $\sim$ 3.5$^{o}$, this 
region is above the main Galactic plane, and the distance estimates to it vary 
from 300 to 900 pc \citep[and references therein]{radhakrishnan72,shu12,val87,rod08}. 
In the present work, we have adopted a distance of 500 pc from 
\citet{radhakrishnan72,shu12}. The central region of the W40 molecular cloud 
and H~{\sc ii} region is known to host an embedded cluster of young stars, 
with a significant population of early-type sources \citep{smi85}, with the 
earliest spectral type of about O9.5 \citep{shu12}. This cluster of early-type 
sources also partially reveals itself as a cluster of compact radio sources, 
coexisting with other compact radio sources which are classified as candidate 
ultra-compact H~{\sc ii} regions and radio variable sources \citep{rod10}. 
Together, these observations show that the W40 molecular cloud/H~{\sc ii} 
region is one of the few nearby regions with active star formation away from 
the Galactic plane, hosting an embedded cluster including high-mass stars, 
and thus representing a template laboratory to investigate the process of 
star formation. 

Recent observations of the larger Serpens-Aquila Rift in the (sub)millimeter 
wavelengths include this region, and have produced a systematic, unbiased 
sample of starless as well as protostellar dense cores within this cloud 
\citep{bon10,mau11}. Also, high resolution X-ray observation using 
\textit{Chandra} has revealed the near-complete census of young stellar 
population within the embedded cluster \citep{kuh10}. Low frequency radio 
observations show newer compact sources in this region \citep{pir12}. 

Despite the multitude of individual studies of the W40 molecular cloud, 
embedded cluster and H~{\sc ii} region, at wavelengths ranging from X-rays 
to radio, an analysis of the overall star formation scenario in W40 is 
pending. In this paper, the aim is to fill this lacuna by using the vast 
archival dataset together with new NIR and radio observations.

In Section \ref{section_Observations-and-Analysis}, we present the observations 
(archival or otherwise) and data reduction procedures. We discuss the results 
pertaining to stellar sources (classification and their analysis), the morphology, 
and the physical parameters of the region from various datasets in 
Sections \ref{section_YSOs}, \ref{section_Morphology}, and 
\ref{section_PhysicalParameters}, respectively; followed by a discussion on the 
possible star formation scenario in Section \ref{section_Star-formation-scenario}. 
The main conclusions are presented in Section \ref{section_Conclusions}. 

~\\
~\\
~\\
~\\
~\\
\section{Observations and Data Analysis}  
\label{section_Observations-and-Analysis} 

\subsection{Near Infrared Photometry} 
\label{section_NIRphotometry}
NIR photometric imaging observations were carried out using the Wide Field Camera 
(WFCAM; \citealt*[]{cas07}) mounted on the 3.8\,m United Kingdom Infrared Telescope 
(UKIRT). WFCAM contains four Rockwell Hawaii-II (HgCdTe 2048 $\times$ 2048 pixel) 
arrays spaced by 94\% in the focal plane.  With a pixel scale of 0.4\arcsec\,, 
the FoV of each array is $\sim$ 13.7\arcmin. \textit{JHK} band data were obtained 
in the service mode during the period 2009 December 15-17. Observations at four 
positions are required to observe a contiguous ($\sim$ 0.75 deg$^2$) square field 
- called a WFCAM \textquoteleft tile\textquoteright\,. 
Observation at each of these positions was taken using a nine point jitter pattern, 
and a 2$\times$2 microstepping pattern, with an exposure time of 5\,s each. The 
seeing was in the range 1.3\arcsec\,-1.8\arcsec\,. We use the {\sc starlink} 
software to do the initial processing, followed by mosaic making using the routine 
{\sc makemos} in a manner similar to \citet{davis07} (see this paper for details 
of WFCAM processing). 

The SExtractor software \citep{bert96} was used to detect all the stellar sources 
in the region. Source detection and deblending is highly effective with this 
software. A detection threshold of 3, and a deblend contrast factor of 32 was used 
for our purposes. Subsequent to first pass of detection, the detection catalogue 
was filtered using the full-width-at-half-maxima (FWHM), stellarness parameter, 
and ellipticity to remove spurious detections. Finally, some leftover spurious 
detections were removed manually to create a reliable detection catalogue. 
Aperture photometry for the detected sources was performed using the {\sc phot} 
routine in the IRAF {\sc apphot} package. The aperture radius was set to 3 pixels 
(= 1.2\arcsec), and the sky annulus was located 9\,pixels away - with a width of
7\,pixels. Sources without any centering error (cier), sky fitting error (sier) 
and photometric error (pier) were retained for further work.

Instrumental magnitudes are in the WFCAM native system (the Mauna Kea Consortium 
Filter System or MKO system). Absolute photometric calibration of the 
\textit{J, H,} and \textit{K} photometry was done using the Two Micron All Sky 
Survey (2MASS) stars in the FoV as explained below. We obtained 2MASS data with 
the constraints of ``phqual = AAA'' and ``ccflag = 000'' for all the stars in the 
FoV. This ensures detections which have the highest photometric quality without 
neighbouring source confusion in all the three bands. We first constructed a 
\textit{J-H/H-K} colour-colour (CC) diagram using this data to identify only the 
main sequence (MS) stars in the region. The MS stars obtained using the CC diagram 
were then used to carry out the photometric calibration of the WFCAM data. The 
2MASS data was converted to WFCAM system using the transformation equations given 
by \citet{hod09}. WFCAM data saturates for magnitudes brighter than 11.5\,mag in 
all the three bands, and these saturated points were replaced by 2MASS magnitudes. 

Completeness limit tests were carried out for all the three bands by choosing 
subimages of size 5\arcmin $\times$ 5\arcmin\ located in the cloud region and 
field region. The 90\% completeness limits for the \textit{J, H}, and \textit{K} 
band data are 19, 18, and 17 mag - respectively - for the field region; and 18, 
17, and 16 mag - respectively - for the cloud region; giving the mean as 18.5, 
17.5, and 16.5 \,mag, respectively. These early observations of WFCAM have small
differences in background matching between multiple chips and tiles, primarily 
resulting from the presence of large scale nebula contaminating the jittered frames. 
Therefore, the \textit{J, H}, and \textit{K} catalogs were clipped at 18, 17 and 
16 mag limits for NIR CC, NIR colour-magnitude (CM), and extinction map analyses 
to obtain conservative and robust classifications of both young stellar objects 
(YSOs) and extinction measurements. Since all NIR magnitudes had low errors 
associated with them, they were not filtered by any error criteria. The magnitudes 
of bright sources which were saturated in the WFCAM image, but had good quality 
(\textquotedblleft phqual=AAA\textquotedblright\,) detections in 2MASS, were taken 
from the 2MASS catalogue after converting them to the WFCAM system using \citet{hod09}.

\subsection{\textit{Spitzer} Photometry} 
\label{section_SpitzerPhotometry} 
The \textit{Spitzer} archival data for the Gould belt survey of 
\citet[PID : 30574]{all06} in all the four IRAC bands - 3.6 (Ch1), 4.5 (Ch2), 
5.8 (Ch3), and 8.0 (Ch4) $\mu$m was obtained using the \textit{Spitzer} Heritage 
Archive. Final image mosaics were created with the MOPEX data reduction software, 
by using the corrected basic calibrated data (cbcd) (S18.5 processed version), 
imask (bimsk), and uncertainty (cbunc) files. Mosaic images in all the four IRAC 
bands were created with a final pixel scale of 0.6\arcsec\,. 

Aperture photometry was carried out for the four IRAC bands. We first used SExtractor 
for source detection as described above (see Section \ref{section_NIRphotometry}) 
and filtered to build a complete detection catalogue. The detection catalogue was 
used as input to the {\sc phot} task in IRAF to obtain the aperture photometry of 
IRAC images.  The aperture size, radius of the inner annulus, and radius of the 
outer annulus were taken as 2.4$^{''}$, 2.4$^{''}$, and 7.2$^{''}$, respectively. 
The zero point magnitudes (with aperture corrections applied) were taken as 18.593, 
18.090, 17.484, and 16.700 for the four IRAC bands, respectively; calculated using 
the IRAC Instrument Handbook (Version 2.0.2, March 2012). 

Average completeness limits were determined using the peak of the luminosity 
function for the four bands. As is elucidated in the \textquotedblleft 
Final Delivery Document For IRAC and MIPS data\textquotedblright\,\footnote
{http://peggysue.as.utexas.edu/SIRTF/} \footnote{http://irsa.ipac.caltech.edu/data/SPITZER/C2D/} 
(see its Section 3.4.2), for c2d legacy project \citep{evans03}, the peak of the 
observed luminosity function can be assumed to estimate the 90\% completeness 
limit. We constructed the luminosity functions with 0.5 mag bin size for each band, 
and the peak (and thus the approximate 90\% completeness limit) was found to be 
at $\sim$ 14, 14, 11.5, and 11.5 for Ch1, Ch2, Ch3, and Ch4, respectively. 

The south-east region of the wide-field image where the extinction is low was the 
only region where both WFCAM and \textit{Spitzer} had a high density of sources 
causing source crowding problems. However, as mentioned in above sections, 
double-care was exercised in filtering first at the level of SExtractor and then 
at photometry allowing us to be quite certain that the detected sources were all 
real and uncontaminated. Finally, the cross-matching makes it even more robust.

\subsection{Cross-matching the Photometric Catalogues}
\label{section_CrossMatching} 
The number of source detections in individual photometric catalogues of each band 
are different due to respective band extinctions, point spread function (psf), 
sensitivity limits and area of coverage. Different cross-matched catalogues were 
produced for different analyses.  The final analysis is done on an area of size 
$\sim$ 34.33$^{'} \times$ 40.00$^{'}$, centered at 
$\alpha_{2000} \sim$ 18$^{h}$31$^{m}$41.23$^{s}$, 
$\delta_{2000} \sim$ -02$^{o}$06$^{'}$29.21$^{''}$. We obtained the following 
cross-matched and merged catalogues for various analyses : 
\begin{itemize}
\item 
\textit{Ch1-Ch2-Ch3-Ch4 catalogue :}
The IRAC Ch1, Ch2, Ch3 and Ch4 band catalogues were cross-matched (within 0.6\arcsec\,) 
with the constraint that all individual band errors are $\leq$ 0.15 mag to produce 
an all-IRAC catalogue of 1874 sources. This was used in the classification of YSOs. 
\item 
\textit{H-K-Ch1-Ch2 catalogue :} 
Not all sources are detected in IRAC Ch3 and Ch4 bands, but have good quality 
detections at \textit{H} and \textit{K} wavelengths. For classification of such 
sources into different evolutionary stages, we obtain a cross-matched catalogue 
of \textit{H}, \textit{K}, Ch1, and Ch2 magnitudes (NIR \textit{H} and \textit{K} 
catalogues were matched with 0.4\arcsec\, matching radius, Ch1 and Ch2 catalogues 
were matched with 0.6\arcsec\, matching radius; followed by matching the resultant 
NIR and \textit{Spitzer} catalogues with 0.6\arcsec\, radius) with good quality 
detections in IRAC Ch1 and Ch2 bands, i.e. sources with IRAC Ch1 and Ch2 magnitude 
errors $\leq$ 0.15 mag. The resulting catalogue contains 8585 cross-matched sources.  
\item 
\textit{J-H-K catalogue :} 
Next, we obtained a cross-matched (within 0.4$^{''}$) \textit{J-H-K} catalogue 
which was further clipped by conservative \textit{J, H}, and \textit{K} completeness 
limits 18, 17, and 16 mag as elucidated in Section \ref{section_NIRphotometry}. 
This filtered NIR catalogue containing 10990 sources was used for the identification 
of YSOs and to produce the extinction map (see Section \ref{section_Extinction}) 
of the region.  
\item 
\textit{H-K catalogue :} 
Many embedded NIR sources are detected only at \textit{H} and \textit{K} wavelengths. 
For identification of additional YSOs from amongst these, cross-matched NIR sources 
(within 0.4\arcsec\,) with detections in only \textit{H} and \textit{K} bands 
(clipped at \textit{H} and \textit{K} conservative completeness limits of 17 and 
16, repectively) were obtained. 
\end{itemize}

The above described catalogues are not mutually exclusive. Respective 
YSOs were identified using these in Section \ref{section_Identification}.

\subsection{H$_2$ narrow-band imaging} 
\label{section_H2NarrowBand} 
Narrow-band imaging in the 2.12 $\mu$m H$_2$ (v=1--0) S(1) filter and
broad band imaging in the \textit{K}$^\prime$ for continuum subtraction 
was obtained for a 19\arcmin$\times$14\arcmin\, region centered on the 
W40 embedded cluster. These data were obtained with the Calar-Alto 3.5\,m 
telescope during the night of 2001 June 13 using the Omega Prime camera. 
The camera used an HgCdTe detector with 1024$\times$1024 pixels with a 
plate scale of 0.39\arcsec\, pixel$^{-1}$ resulting in a 
6.75\arcmin$\times$6.75\arcmin\, FoV. Observations were carried out using 
a 9 point jitter pattern. The exposure time was 30\,s and 3\,s - per 
jitter pointing - in the NB2122 and \textit{K}$^\prime$ filters 
respectively, resulting in total integration times of 180\,s and 18\,s 
per pixel, respectively. Standard image reduction of dark subtraction, 
flat-fielding and sky-subtraction was carried out for each pointing. 
The \textit{K}$^\prime$ images were psf matched and scaled to the H$_2$ 
narrow-band image to perform continuum subtraction.

\subsection{\textit{Herschel} Archival Data} 
\label{section_HerschelData} 
\textit{Herschel} Space Observatory, a 3.5\,m telescope, was launched to carry 
out observations in the wavelength range 51-670 $\mu$m using the instruments 
Photodetector Array Camera and Spectrometer (PACS), SPIRE, and HIFI \citep{pil10}. 
We obtained the publicly available level2\_5 processed archival data for PACS 
100 $\mu$m (3.2\arcsec\,pixel$^{-1}$) and 160 $\mu$m (6.4\arcsec\,pixel$^{-1}$) 
bands - obtained as part of the Proposal ID:KGBT\_pandre\_1. 
The \textquotedblleft MADmap\textquotedblright \citep{can10,was07} processed 
products are used, as opposed to \textquotedblleft PhotProject\textquotedblright\, 
processing, since our aim is to analyse extended features in the region. 
A detailed overview of procedures involved, data reduction pipeline, and data 
products is available on the \textit{Herschel} website 
\footnote{http://herschel.esac.esa.int/Data\_Processing.shtml}
\footnote{http://herschel.esac.esa.int/Data\_Products.shtml}
We use the images to examine the filamentary structures vis-a-vis the distribution 
of YSOs and radio continuum emission.

\begin{deluxetable}{c c c}
\tablecolumns{3}
\tablewidth{0pt}
\tabletypesize{\tiny}
\tablecaption{Details of GMRT Observations}
\tablehead{
\colhead{} & \colhead{1280 MHz} & \colhead{610 MHz}}
\startdata
Date of Obs. & 2011 Nov. 15 & 2011 Nov. 18 \\
Phase Center & $\alpha_{2000}$=18$^h$31$^m$15.75$^s$ & $\alpha_{2000}$=18$^h$31$^m$15.75$^s$ \\
             & $\delta_{2000}$=-02$^o$06$^{'}$49.30$^{''}$ & $\delta_{2000}$=-02$^o$06$^{'}$49.30$^{''}$ \\
Flux Calibrator & 3C286, 3C48 & 3C286, 3C48 \\
Phase Calibrator & 1822-096 & 1822-096 \\
Cont. Bandwidth & 32 MHz & 32 MHz \\
Primary Beam & 24$^{'}$ & 43$^{'}$ \\
Resolution of maps  &  & \\
used for fitting & 45.84$^{''}\times$44.40$^{''}$ &  44.11$^{''}\times$40.67$^{''}$ \\
Peak Flux Density & 0.45 Jy/beam & 0.35 Jy/beam \\
rms noise & 6.34 mJy/beam & 5.52 mJy/beam \\
Total Integrated  & & \\ 
Flux Density & & \\
within 4$\sigma$ area & 7.55 Jy & 4.07 Jy \\ 
Integrated Flux  & & \\ 
Density within   & & \\
50\arcsec\, radius of & & \\
IRS\,5 radio peak  & 109 mJy & 121 mJy \\ 
\enddata
\label{table_GMRTObservation}
\end{deluxetable}

\subsection{Radio Continuum Observations}

New radio continuum observations are obtained using the Giant
Metre-wave Radio Telescope (GMRT) in two bands centered at 1280 MHz 
and 610 MHz. Data was obtained on 2011 November 15, and 2011 November 
18 (Project ID 21\_015), respectively, the details of which are given 
in Table \ref{table_GMRTObservation}. The GMRT array consists of 30 
antennae located in an approximately Y shaped configuration. Each
antenna has a diameter of 45\,m. A central region of about 1\,km
$\times$ 1\,km consists of 12 (randomly distributed) antennae, while
the rest are distributed along three radial arms (6 along each arm)
upto a distance of $\sim$ 14\,km. It results in a baseline of
$\sim$ 100\,m--25\,km, providing sensitivity to features $\lesssim$
8$^{'}$ and 17$^{'}$ in the 1280 MHz and 610 MHz bands,
respectively. Details about the GMRT array can be found in \citet{swa91}.

Data reduction was carried out with the AIPS software. Bad data were
flagged using AIPS tasks {}``TVFLG'' and {}``UVFLG''. Multiple rounds
of flagging and calibration were done to improve the calibration
successively. Calibrated object data (W40) was subsequently split from
the raw file (which contains phase and flux calibrator data also),
following which low resolution ($\sim$ 45$^{''} \times$ 45$^{''}$)
images were finally made using facet imaging with the help of the task
{}``IMAGR''. A few rounds of (phase) self-calibration were carried out
to remove the ionospheric phase distortion effects.

\begin{figure*}
\centering
\subfigure
{
\includegraphics[trim={1cm 6.5cm 1cm 9.5cm}, clip, scale=0.45]{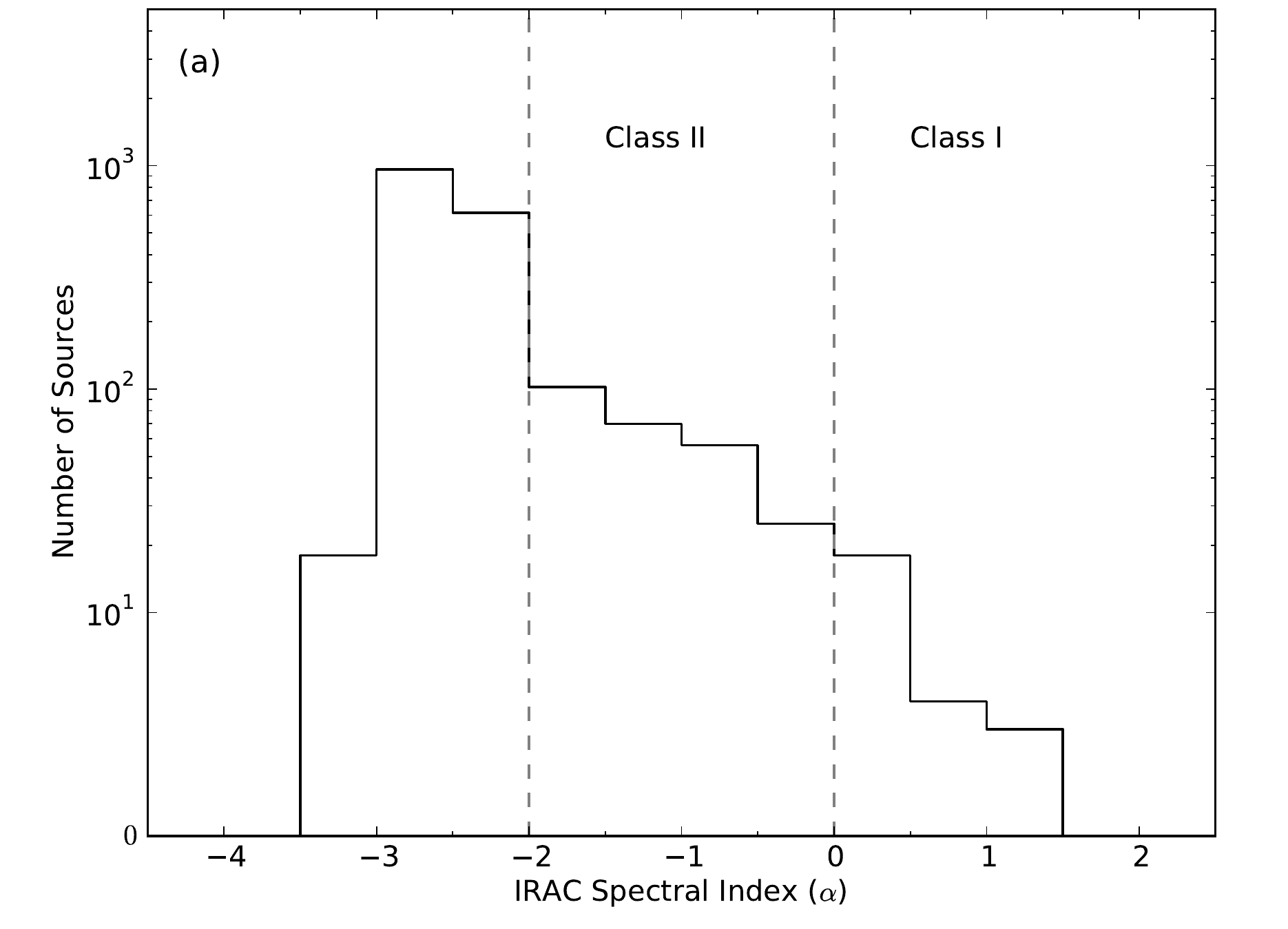}
\label{fig_SpectralIndex}
}
\subfigure
{
\includegraphics[trim={1cm 7cm 1cm 9cm}, clip, scale=0.45]{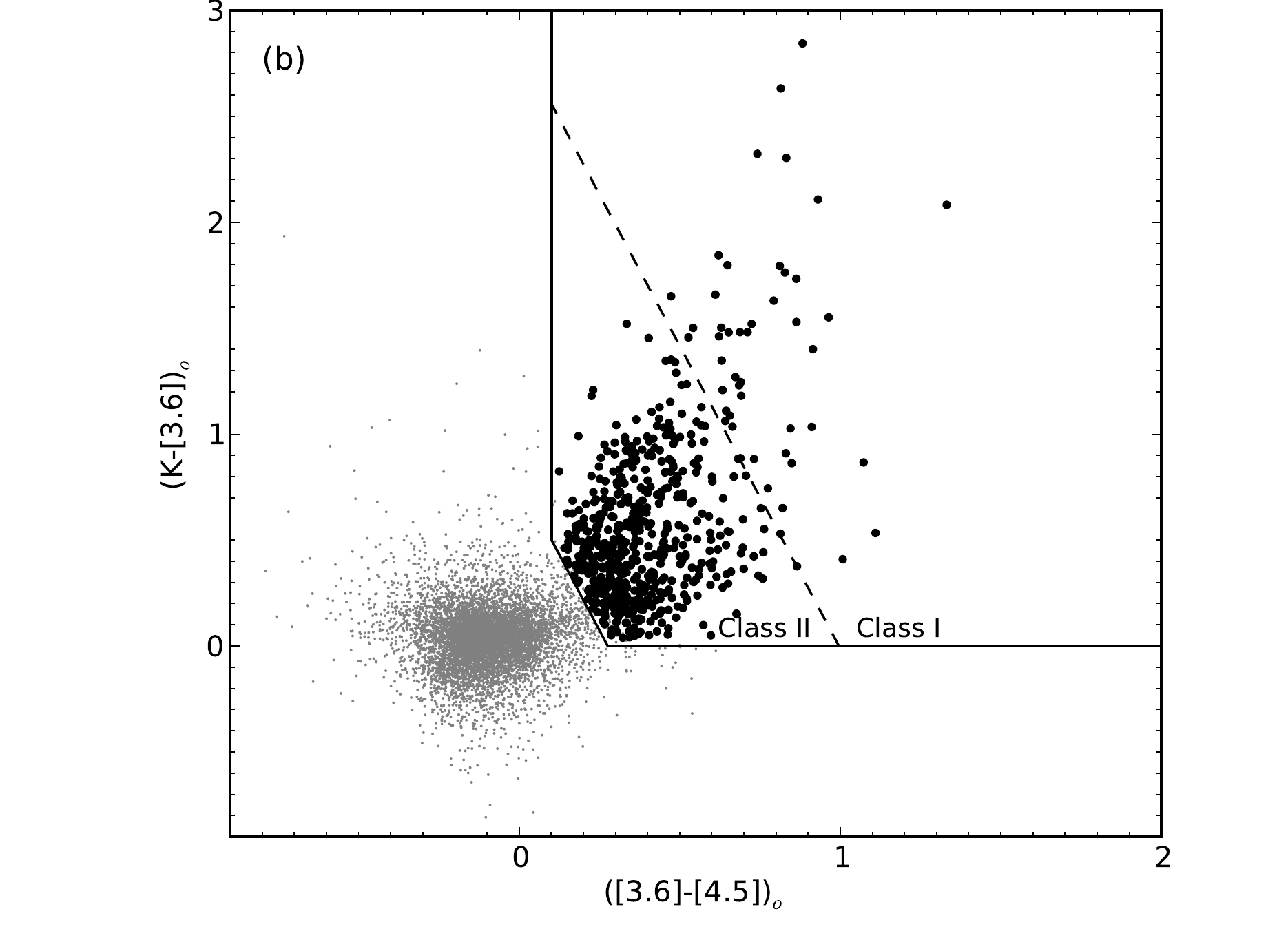}
\label{fig_HKCh1Ch2_CCD} 
}
\subfigure
{
\includegraphics[trim={1cm 7cm 1cm 8cm}, clip, scale=0.45]{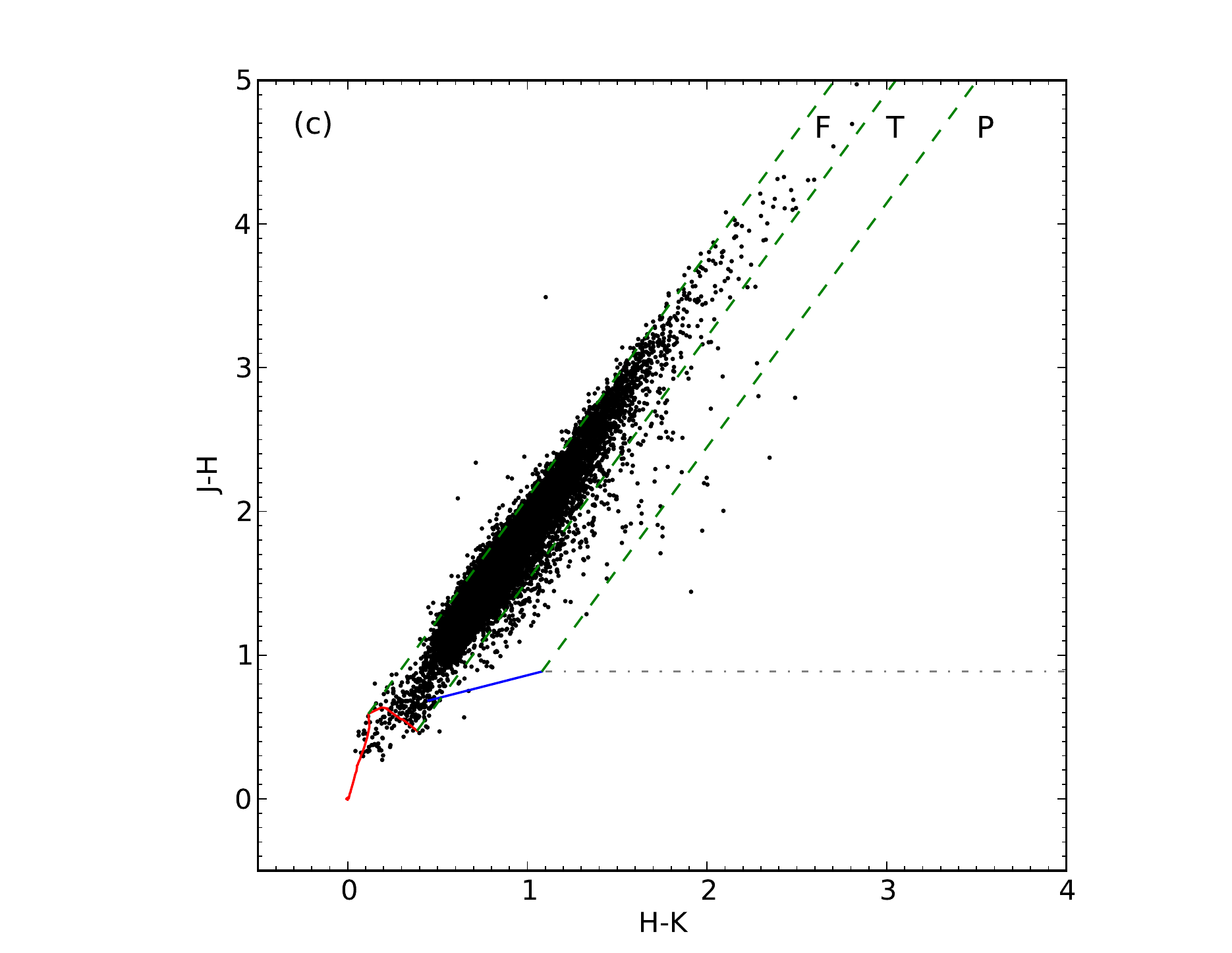}
\label{fig_JHK_CCD} 
}
\subfigure
{
\includegraphics[trim={1cm 7cm 1cm 8cm}, clip, scale=0.45]{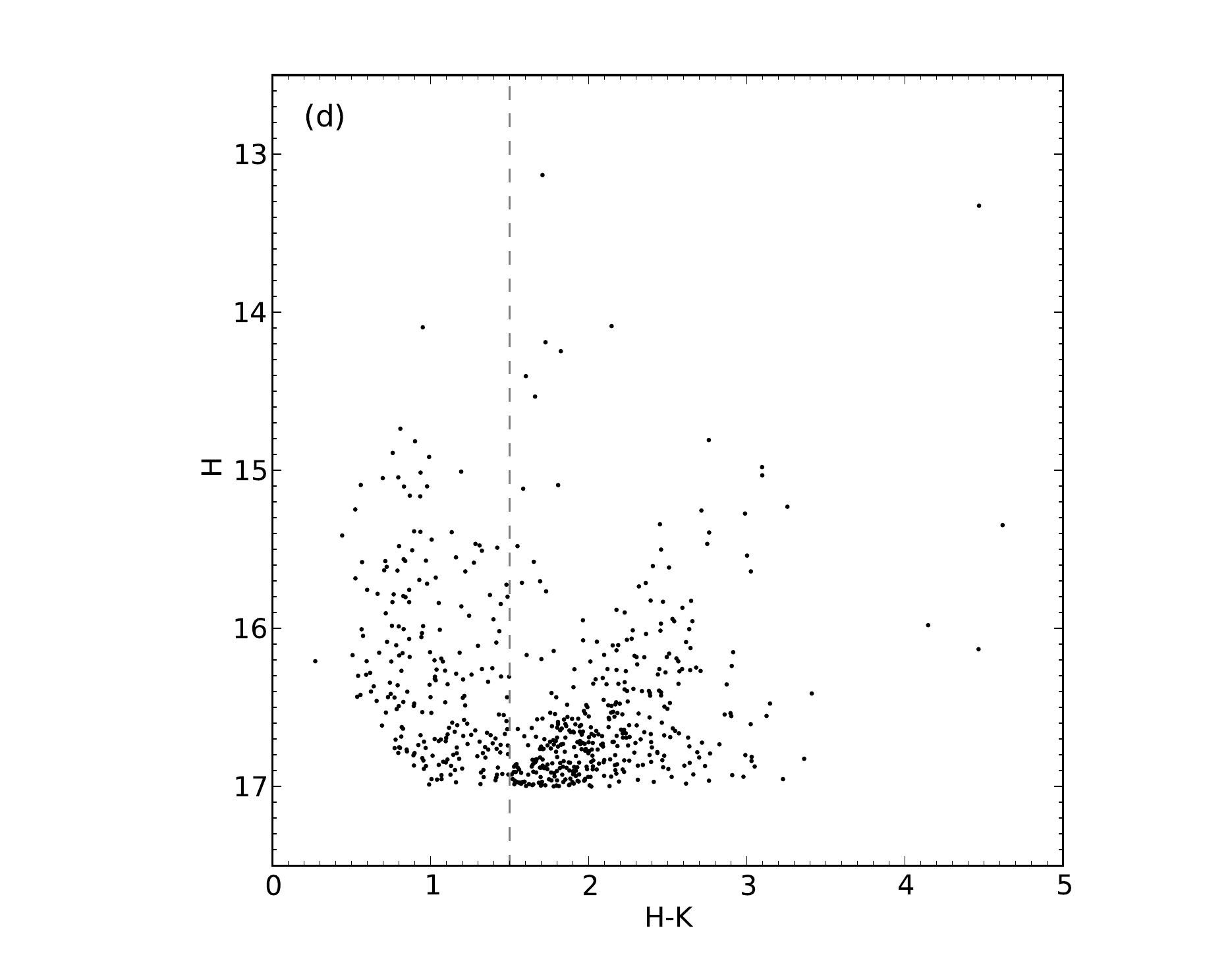}
\label{fig_HK_CMD} 
}
\caption{(a) Histogram of IRAC spectral index of sources. The boundary of Class I 
and Class II sources has been marked with vertical grey dashed lines. 
(b) NIR/IRAC CC plot showing the location of Class I and Class II sources, with 
areas demarcated (by black dashed and solid lines) using the procedure of \citet{gut09}. 
The filled black circles mark the sources identified. 
(c) \textit{J-H/H-K} CC diagram for the region. Solid blue line is the CTTS locus 
from \citet{yasui08}. Red curve marks the locus of the dwarfs from \citet{hewett06}. 
Dashed green lines are the reddening vectors, drawn (starting from the turning 
point, and the low-mass end of dwarf locus; and upper end of CTTS locus) using the 
reddening laws from \citet{rieke85}. Three regions - \textquoteleft F\textquoteright\,, 
\textquoteleft T\textquoteright\,, and \textquoteleft P\textquoteright\, mark the 
location of different classes of YSOs. The horizontal grey dot-dashed line has 
been drawn at $J-H = 0.89$. All points are in MKO system. 
(d) \textit{H/H-K} CM diagram for the NIR sources with detections in only \textit{H} 
and \textit{K} bands. The grey vertical line (at \textit{H-K} = 1.5) marks the 
cut-off upwards of which YSOs have been selected.} 
\label{fig_YSO-plots} 
\end{figure*}

\section{Young Stellar Population}
\label{section_YSOs}

\subsection{Identification of YSOs}
\label{section_Identification} 

The YSOs were identified using sources detected in IRAC bands first, followed by 
using the catalogue of sources with detections in NIR and IRAC bands, and lastly 
using the NIR catalogue of sources. Following is an elucidation of the set of 
steps followed : 
\begin{itemize}
\item 
Identification of YSOs from the IRAC data was accomplished by computing the IRAC 
spectral index for the sources with detections in all four IRAC bands (i.e sources 
from \textit{Ch1-Ch2-Ch3-Ch4} catalogue; see Section \ref{section_CrossMatching}). 
The IRAC spectral index 
\citep[$\alpha_{IRAC} = d\,\log(\lambda F_{\lambda})/d\,\log(\lambda)$; ][]{lada87}, 
for each source, was calculated using linear regression. Thereafter, using the 
spectral index limits as given in \citet{chavarria08} (Class I : 0 $< \alpha_{IRAC}$, 
and Class II : -2 $\leq\alpha_{IRAC}\leq$ 0), sources were classified into 
Class I and Class II categories (see Figure \ref{fig_SpectralIndex}). 
\item
\textit{H-K-Ch1-Ch2} cross-matched catalogue (see 
Section \ref{section_CrossMatching}) was used for the identification of 
additional Class I and Class II sources using the classification scheme 
of \citet[see their Appendix A - Phase 2]{gut09,gut10}. For this, the 
dereddened (K-[3.6]) and ([3.6]-[4.5]) colours, calculated using the colour 
excess ratios from \citet{fla07}, were used. As is stated in \citet{gut09}, 
to filter out possible dim extragalactic contaminants, we also applied the 
limits on dereddened Ch1 band magnitude to be $\leq$ 15 and $\leq$ 14.5 for 
Class I and Class II categories, respectively. We used the average 
extinction value estimated in Section \ref{section_Extinction}) 
\citep[converted to A$_K$ using][]{rieke85} to calculate the extinction for 
Ch1 band in this region, and thus intrinsic Ch1 magnitudes, to impose this 
condition. The reddening law from \citet{fla07} for Serpens (since it is 
closest to our region) was used for this purpose. Figure \ref{fig_HKCh1Ch2_CCD} 
shows the location of these Class I and Class II sources - which have been 
marked as filled black circles - on the 
(\textit{K}-[3.6])$_o$ vs. ([3.6]-[4.5])$_o$ CC diagram.

\begin{deluxetable*}{c c c c c c c c c c}
\tablecolumns{11}
\tablewidth{0pt}
\renewcommand{\arraystretch}{1}
\tabletypesize{\tiny}
\tablecaption{YSOs identified in the FoV \label{table_yso}} 
\tablehead{ \colhead{RA}    & \colhead{Dec}   & \colhead{\textit{J}}      & \colhead{\textit{H}}     & \colhead{\textit{K}}    & \colhead{[3.6]}          & \colhead{[4.5]} & \colhead{[5.8]}     & \colhead{[8.0]}   & \colhead{Class}  \\   
\colhead{(J2000)} & \colhead{(J2000)} & \colhead{(mag)}   & \colhead{(mag)} & \colhead{(mag)}         & \colhead{(mag)} & \colhead{(mag)} & \colhead{(mag)}  & \colhead{(mag)}    &  \colhead{}   }    
\startdata
277.635620    &   -2.305891    &   14.637  $\pm$  0.038    &   11.474  $\pm$  0.025    &    9.499  $\pm$  0.025    &    8.417  $\pm$  0.002    &    8.249  $\pm$  0.002    &    7.810  $\pm$  0.004    &    7.860  $\pm$  0.004    &   Class2     \\
277.638031    &   -2.118266    &          \nodata          &          \nodata          &          \nodata          &   11.351  $\pm$  0.006    &   10.719  $\pm$  0.006    &   10.156  $\pm$  0.016    &    9.422  $\pm$  0.016    &   Class2     \\
277.638184    &   -2.349572    &   14.582  $\pm$  0.048    &   11.551  $\pm$  0.025    &    9.273  $\pm$  0.020    &    7.709  $\pm$  0.001    &    6.973  $\pm$  0.001    &    6.402  $\pm$  0.002    &    6.179  $\pm$  0.001    &   Class2     \\
277.638580    &   -2.190570    &          \nodata          &          \nodata          &          \nodata          &   10.309  $\pm$  0.004    &    9.623  $\pm$  0.003    &    8.998  $\pm$  0.009    &    8.210  $\pm$  0.004    &   Class2     \\
277.641449    &   -1.932326    &          \nodata          &   14.189  $\pm$  0.002    &   12.462  $\pm$  0.001    &          \nodata          &          \nodata          &          \nodata          &          \nodata          &   Class2     \\
\enddata
\vspace{-0.4cm}
\tablecomments{Only a portion of the table is shown here. The complete table is available in electronic form as part of the online 
material.} 
\end{deluxetable*}

\item 
In the next step, the \textit{J-H-K} photometric catalogue was used as follows. 
For the sources in this catalogue, a \textit{J-H/H-K} CC diagram was constructed. 
Figure \ref{fig_JHK_CCD} shows the NIR \textit{J-H/H-K} CC diagram. The Classical 
T-Tauri Star (CTTS) locus (blue line) was taken from \citet{yasui08} 
\citep[which is derived from][]{meyer97}, while the locus for dwarfs (red curve) 
in MKO system was taken from \citet{hewett06}. Similar to \citet{yasui08}, the 
reddening laws of \citet{rieke85} (A$_J$/A$_V$ = 0.282, A$_H$/A$_V$ = 0.175, and 
A$_K$/A$_V$ = 0.112) were used. The dashed green lines show the reddening vectors. 
The sources which were located in the \textquoteleft T\textquoteright\, and 
\textquoteleft P\textquoteright\, regions - similar to the method of 
\citet{ojh04a,ojh04b}, \citet{lada92} - were taken as NIR Class II/III sources. 
There may be an overlap between the Herbig Ae/Be stars and Class II sources in 
this \textquoteleft T\textquoteright\, region \citep{hillenbrand92}. To decrease 
any contamination in the \textquoteleft P\textquoteright\, region, we took only 
those sources which were above $J-H=$ 0.89 threshold. 
\item 
Additionally, those NIR sources detected only in the \textit{H} and \textit{K} 
bands (i.e. using the \textit{H-K} catalogue from Section \ref{section_CrossMatching}) 
were also used to identify YSOs using the \textit{H/H-K} CM diagram 
(Figure \ref{fig_HK_CMD}). From the \textit{H/H-K} CM diagram, only the sources 
with an IR excess upwards of \textit{H-K} = 1.5 were selected. This cut-off was 
chosen as - in the CM diagram - it marks a clear low density gap between the field 
sources branch and the YSOs' branch. CM diagrams of nearby non-nebular regions 
also show the sources in those regions (i.e. the field sources) to be confined 
below this value. YSOs selected using the NIR photometry mostly fall into the 
Class II/III category. 
\end{itemize}
The YSOs selected from each of the above steps are not mutually exclusive and 
there are overlapping YSOs between various methods. The selected YSOs were hence
merged (with the class of a YSO taken as what it was identified as first in the 
above order of steps) to obtain a final catalog of 1202 YSOs, of which 40 are in 
the Class I category and 1162 in the Class II/III category. 
The central high-mass IRS sources (1A South, 2B, 3A, and 5) were detected as 
Class\,II in the above set of steps, but were expunged from the YSO catalogue 
as they have already been identified as main-sequence sources by \citet{shu12}. 
Table \ref{table_yso} gives the YSOs detected in the FoV. A sample of 
Table \ref{table_yso} is given here, whereas the complete table is available in 
electronic form as part of the online material.

\begin{figure}
\includegraphics[trim={0cm 0cm 0cm 8cm}, clip, scale=0.45]{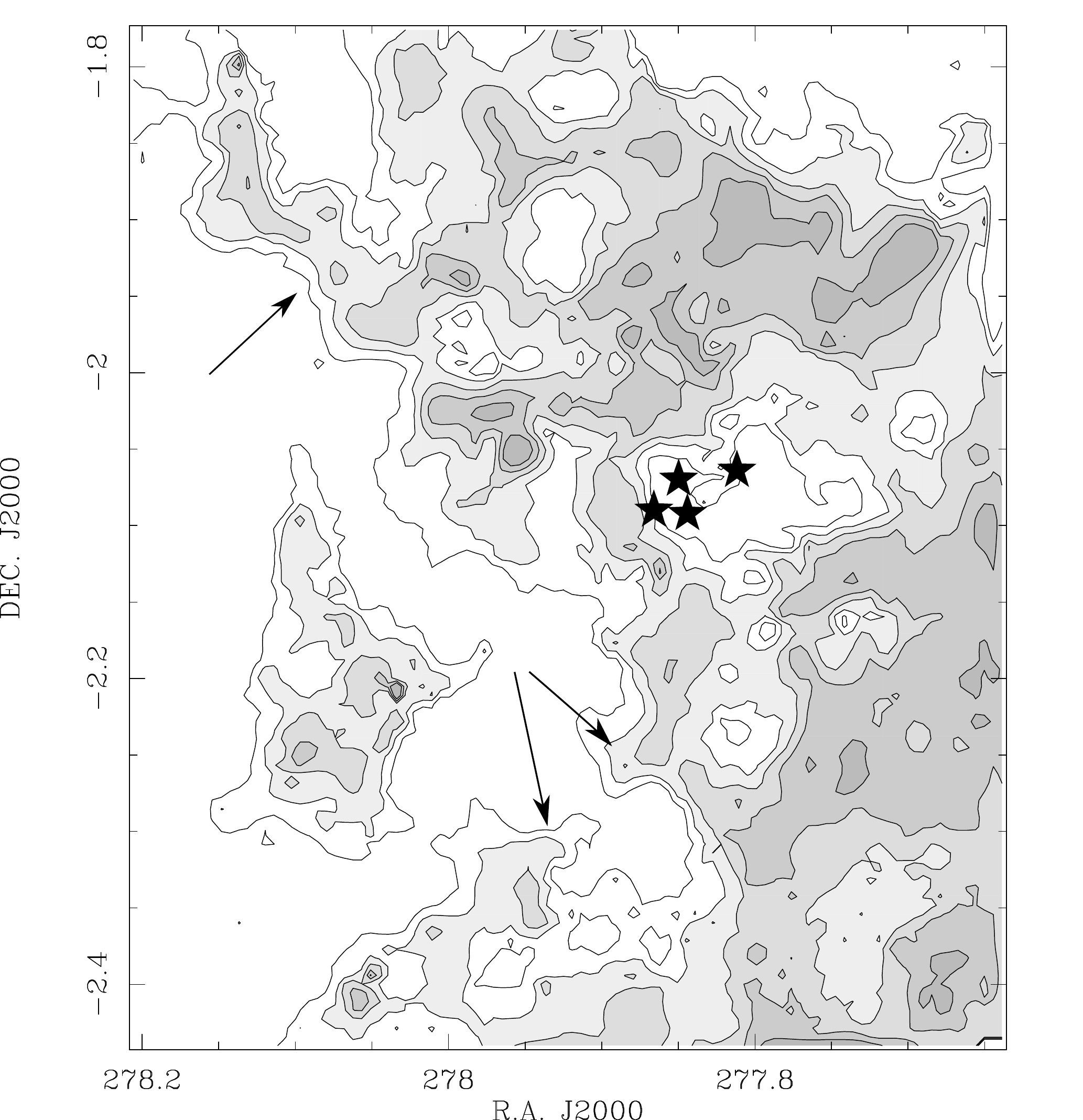}
\caption{Visual extinction map of the region calculated using median value of 20 
nearest neighbours. The (A$_V$) contours are drawn from 12 to 20 mag, with steps 
of 2 mag. The star symbols mark the location of central high-mass stars. Arrows
indicate the locations of filamentary structures}.  
\label{fig_ExtMap}
\end{figure}

\subsection{Extinction in the Region} 
\label{section_Extinction}

The NIR photometry from the new WFCAM data is deeper by two magnitudes
compared to the 2MASS data. The NIR \textit{J-H/H-K} CC diagram of the
region was used to select the reddened MS sources. Sources in the 
\textquoteleft F\textquoteright\, region (see Figure \ref{fig_JHK_CCD}), 
which were not classified as YSOs in Section \ref{section_Identification}, 
were selected for this analysis (a total of 10011 sources used here). 
We estimated the average visual 
extinction in the region by using a method similar to the 
Near-IR-Colour-Excess (NICE) method of \citet{lad94, kai07, ojh04a}. First, 
the \textit{(H-K)} colour excess for each of the selected sources was 
calculated by dereddening them, along the reddening vectors, to the dwarf 
locus which was approximated by a straight line asymptote. A$_V$, for each
source, was subsequently calculated using the reddening laws of 
\citet{rieke85}.

\begin{figure*}[ht]
\centering
\includegraphics[trim={0cm -0.5cm 0cm 16cm}, clip, scale=0.8]{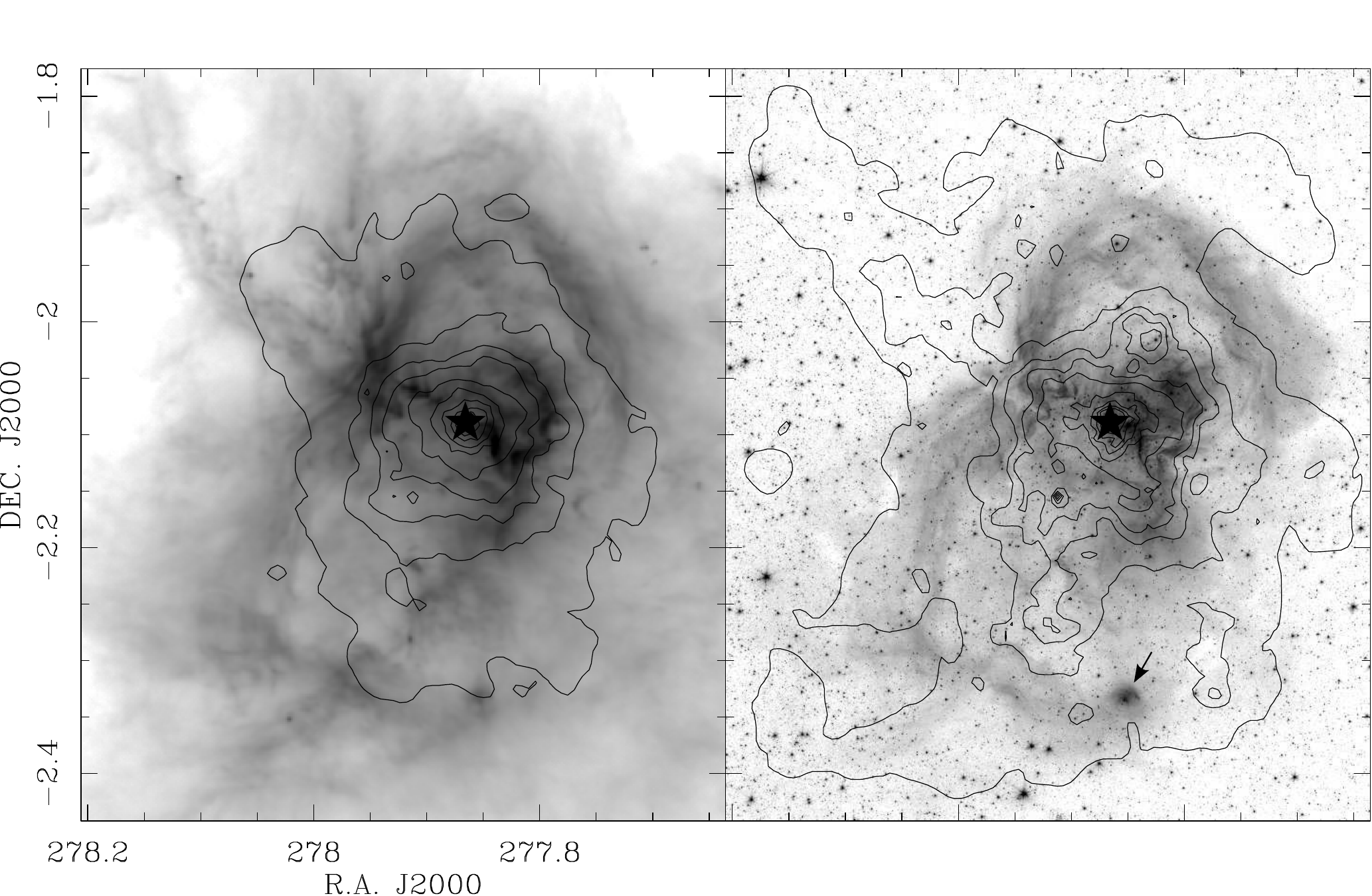}
\caption{Surface Density contours of the region. 
\textit{(left)} Made using 50 nearest neighbours and overlaid on PACS 160 $\mu$m image. 
The contours are drawn at 40, 70, 100, 130, 200, 300, 400, 500, and 600 YSOs pc$^{-2}$. 
\textit{(right)} Made using 20 nearest neighbours and overlaid on IRAC 3.6 $\mu$m image. 
The contours are drawn at 25, 50, 75, 100, 150, 275, 400, 525, and 650 YSOs pc$^{-2}$. 
The arrow marks a prominent compact nebula \textit{(see text)}. 
The star symbol in both the figures marks the location of IRS 1A. } 
\label{fig_SurfaceDensity}
\end{figure*}

The dereddenned extinction values were used to produce an extinction
map of the region as shown in Figure \ref{fig_ExtMap}. In producing 
this extinction map, at each position on the regular grid, a median 
extinction from the 20 valid nearest neighbours was computed. The 
choice of median instead of average value is because median works like 
an outlier rejection. Figure \ref{fig_ExtMap} displays two distinct 
features : 
a) a low extinction cavity around the ionizing stars, and b) higher 
extinction towards west and north. The filamentary structures (whose 
locations have been marked with arrows) are partially traceable (with 
the one along north-east much more distinct than the one towards south). 
Although the filaments are found to be very dense from PACS 160 $\mu$m
emission (see Section \ref{section_FilamentaryStructures}), the 
extinction map fails to reproduce the same result because the NIR 
data is not deep enough to effectively detect many stars behind the 
filament. The depth of the NIR data is nevertheless good enough to 
trace extinctions of the order of A$_V \sim$ 20\,mag, which is the 
maximum value found along the north-eastern filamentary structure  
and northern regions. 
 
The average extinction of the diffuse low density remnant cloud covering 
most of the W40 H~{\sc ii} region is found to be A$_V \sim$ 15\,mag. Note 
that the low extinction cavity around the IRS sources in 
Figure \ref{fig_ExtMap} displays A$_V \le$ 12\,mag which agree well with 
the previous extinction estimates for the dominant IRS sources (A$_{V} 
\sim$ 8-10 mag) in the region \citep{smi85,shu99,rod10}.

\begin{figure*}
\centering
\includegraphics[trim={2cm 9.5cm 1cm 11cm}, clip, scale=1]{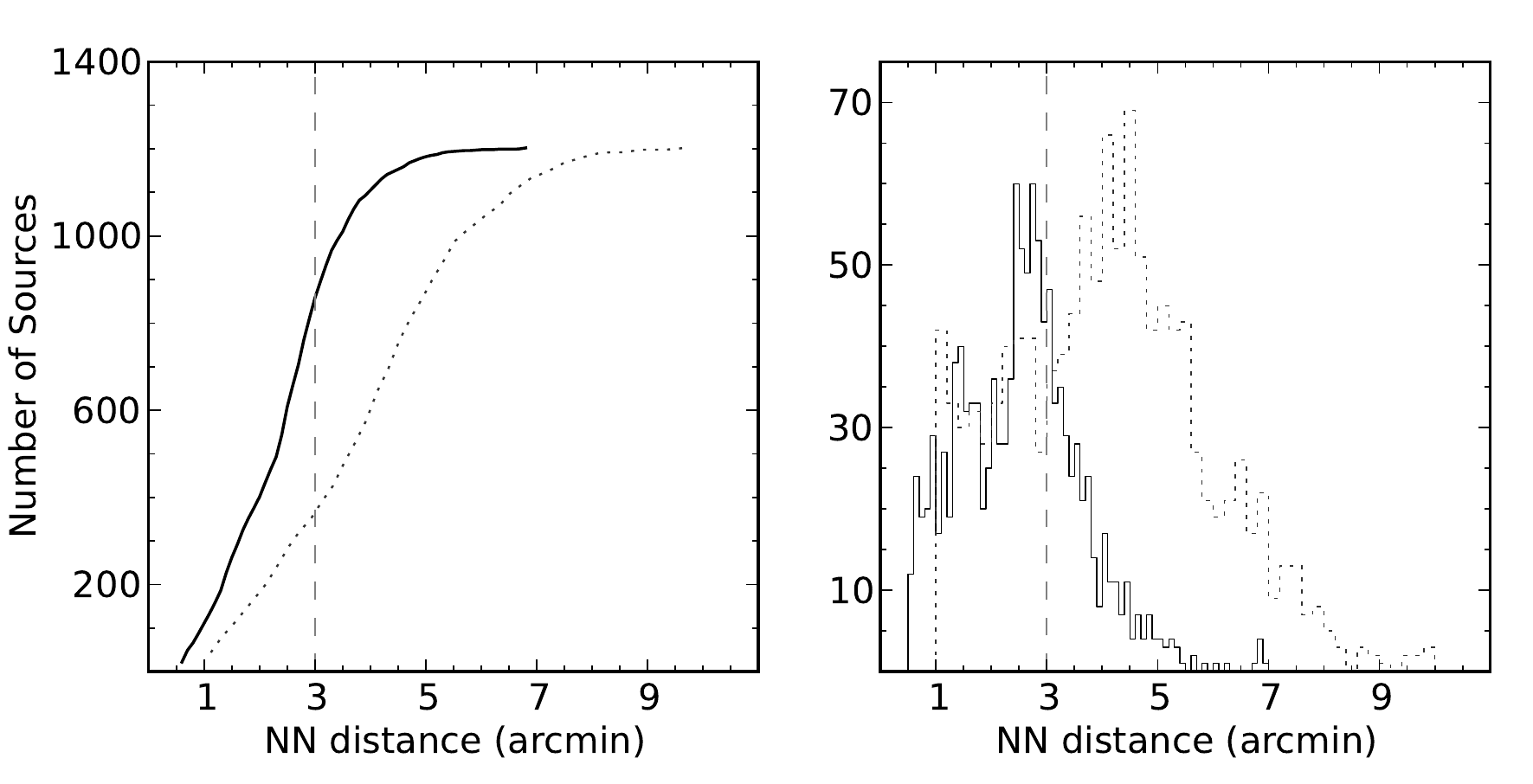}
\caption{\textit{(left)} The frequency polygon for the
cumulative frequency distribution of the nearest neighbour (NN) distances, 
and \textit{(right)} the histogram of the NN distances. The NN distances are 
in arcminutes. The solid black line/curve is for 20 NN distances, while 
the dark-grey dotted line/curve is for 50 NN distances. The binsizes taken 
are 0.10 and 0.20 for 20 NN and 50 NN, respectively. The vertical grey dashed line 
marks the cluster radius as calculated from 50 NN map in 
Figure \ref{fig_SurfaceDensity}.} 
\label{fig_DensityHistograms}
\end{figure*}

\subsection{Surface density} 
\label{section_SurfaceDensity} 

We carried out the YSO surface density analysis of the region using the nearest 
neighbour (NN) method \citep{casertano85, schmeja08, schmeja11}. Figure \ref{fig_SurfaceDensity} 
shows contours for 50 and 20 NN density overlaid on PACS 160 $\mu$m image and 
IRAC 3.6 $\mu$m image, respectively. 
A large value of NN is used to examine large scale structures, while a lower value 
of NN is more sensitive to smaller-scale density variations and is used to trace 
sub-structures. This can be thought of some kind of smoothing process. Here, the 
50 NN density contours are fairly circularly symmetric, and were thus used to 
calculate the cluster radius (as opposed to using 20 NN map). The dense central 
region for 50 NN map was fitted by a Gaussian profile whose FWHM comes out to be 
$\sim$ 3\arcmin\, ($\sim$ 0.44 pc at a distance of 500 pc). There are 170 YSOs 
(from Section \ref{section_Identification}) within this cluster radius. The 20 NN 
density contours (being more sensitive to local density fluctuations) reveal the 
sub-structures of the cluster. The contours mirror the filamentary structures to 
a considerable degree showing that the filaments also contain YSOs, albeit at low 
densities. The centers of both 50 and 20 NN density contour maps are coincident 
with IRS\,1A. 
Figure \ref{fig_DensityHistograms} shows the cumulative frequency distribution 
and the histogram for the 20 NN and 50 NN distances in solid black line/curve 
and dark-grey dotted line/curve, respectively. 
As is expected, each feature (kink, peak, dip, etc) which is observed in 50 NN 
histogram/frequency polygon at, say, \textquotedblleft x\textquotedblright\,  
NN distance, is observed at $\sim$  \textquotedblleft x/2\textquotedblright\,  
NN distance in 20 NN histogram/frequency polygon. The cluster radius has been 
marked in vertical dashed line. The cluster radius seems to be near an inflection 
point in the rising portion of the 50 NN frequency polygon.

\subsubsection{Comparison with other regions} 
\label{section_SurfaceDensityComparison} 

\citet{schmeja08} have carried out an analysis of clusters in nearby star-forming 
regions - \object{Perseus}, \object{Serpens}, and \object{Ophiucus} molecular 
clouds - using \textit{Spitzer} data. The cluster peak for the W40 region 
($\sim$ 650 pc$^{-2}$ for 20 NN) seems to be much higher than the clusters in these 
molecular clouds, except for that in Serpens Core(A) (which is 1045 pc$^{-2}$), 
and comparable to \object{L1688} in Ophiucus (which is 509 pc$^{-2}$); while the 
cluster radius is much smaller (at least a factor of $\sim$ 0.5 or more lesser) 
than all of them, except for that of Ophiucus Centre cluster (which has a radius 
of 0.52 pc). 
From the compilation of \citet{lada03} for different clusters (though it must be 
kept in mind that this is pre-\textit{Spitzer} and mostly just using NIR), we can 
see that among the few clusters at similar distances and with similar cluster 
radii (e.g. \object{L1641N}, \object{L1641C}, \object{MWC\,297}, \object{S\,106}), 
the star count is mostly much lower than for W40 region (which is 170). Even among 
regions at larger distances and similar radii (e.g. \object{01546+6319}, 
\object{02407+6047}, \object{02497+6217}, \object{02541+6208}, \object{IRAS\,06068+2030}, 
\object{IRAS\,06155+2319}, \object{MWC\,137}), the star counts are much lower.

\begin{figure*}
\centering
\includegraphics[trim={1.3cm 7cm 1.5cm 9cm}, clip, scale=0.90]{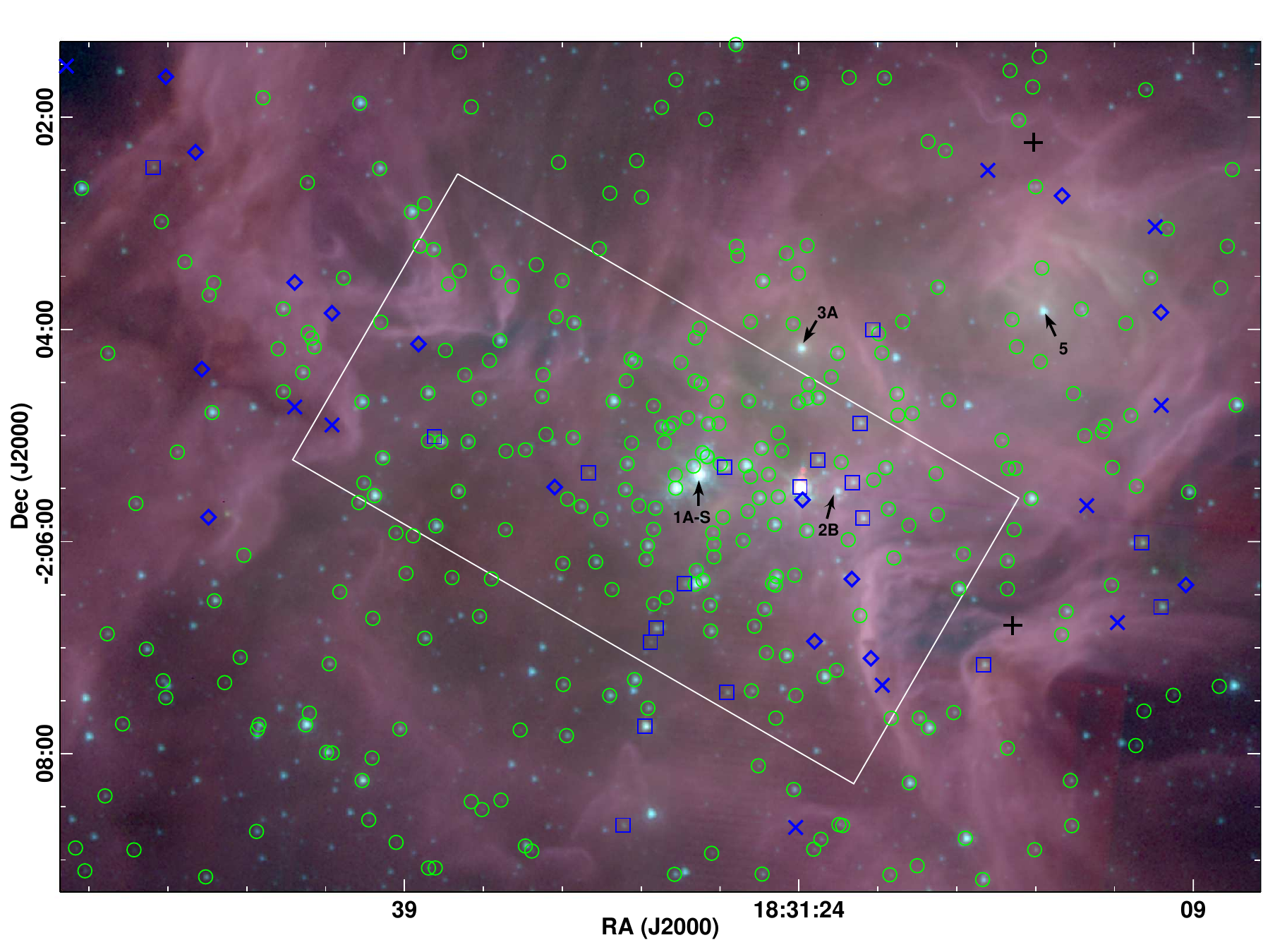}
\caption{Colour composite image of the zoomed-in central region using IRAC 8.0 
$\mu$m (red), 4.5 $\mu$m (green), and 3.6 $\mu$m (blue) images. The high-mass IRS 
sources (1A-South, 2B, 3A, and 5) have been marked by black arrows, while the 
$^{12}$CO(J=1-0) peaks \citep{zei78,smi85} have been marked by black plus symbols. 
Green circles and blue squares denote the Class II/III and Class I sources, 
respectively. Blue diamond and blue cross symbols denote protostars and starless 
cores, respectively, from \citet{mau11}. The white rectangle denotes possible 
\textquotedblleft hub\textquotedblright\, region 
(see Section \ref{section_FilamentaryStructures}).} 
\label{fig_CentralArea}  
\end{figure*}

\subsection{Spatial distribution and various features} 
\label{section_SpatialDistribution} 

The YSOs are not in particular found to be correlated with the lobes of the bipolar 
nebula, suggesting that the bipolar nebula is relatively young, and not enough 
material has accumulated along the edge of the lobes to lead to star formation. 
There appears to be a prominent compact nebula (marked with a black arrow on 
3.6 $\mu$m image in Figure \ref{fig_SurfaceDensity}) located at the edge of the 
southern lobe. The source associated with this compact nebula was not classified 
as any YSO (though there are Class II/III YSOs in its vicinity). It is likely to 
be a reflection nebula illuminated by the embedded star, and seems to have been 
overrun by the expanding southern wall of the bipolar nebula. 

Figure \ref{fig_CentralArea} shows the zoomed-in view of the central region with 
overlaid YSOs. Class I and Class II/III sources have been marked in blue squares 
and green circles, respectively. 
While Class\,II/III sources are distributed throughout the image, the Class\,I 
sources are located mostly along the filamentary structures extending upto beyond 
the cluster radius (details in Section \ref{section_FilamentaryStructures}). 
Additionally, protostars and starless cores from \citet{mau11} have been marked 
in blue diamond symbols and crosses, respectively. We would like to note here that 
none of the Class 0/I sources or starless cores from \citet{mau11} had any YSO 
counterpart from our analysis within 1\arcsec\, matching radius, probably because 
these \citet{mau11} sources are younger (and/or are oriented edge-on) as opposed 
to the YSOs identified in Section \ref{section_Identification} which are Class I 
or Class II/III sources. It is possible that smoothing by \citet{mau11} to 
13\arcsec\, beam - while we are only matching upto 1\arcsec\, here - might be the 
cause of mismatch. Moreover, the classification by \citet{mau11} is using envelope 
mass vs. bolometric luminosity diagram, inducing further uncertainty.  

The high-mass IRS sources have been marked with arrows and labelled. IRS\,1A South 
- the main ionizing source - was found, using CC/CM diagrams, to exhibit an IR 
excess emission. This could be because it is located in the central region 
surrounded by high density material. IRS 2B, 3A, and 5 (other high-mass sources, 
but of comparatively much lower mass than 1A South) were found to have a negative 
SED slope, with the respective spectral indices being $\sim$ -1 for IRS\,2B, 
-0.4 for IRS\,3A, and -1.9 for IRS\,5.

\subsubsection{IRS\,5 nebular region}
\label{section_SpatialDistributionIRS5Nebula} 

Another noticeable feature is the arc-shaped nebula around the IRS\,5 high mass 
source. \citet{shu12} have estimated the spectral type of this source as B1 type. 
The protostars and starless cores from \citet{mau11} seem to be distributed along 
the circumference of this arc-shaped nebula. This probably indicates that material 
has been pushed by the expansion of H~{\sc ii} region of IRS\,5 and collection of 
dense material due to it has led to the formation of these protostellar sources 
and cores. 

\subsubsection{Bright-Rimmed Clouds}
\label{section_SpatialDistributionBRC} 

The zoomed-in central region shown in Figure \ref{fig_CentralArea} with overlaid 
sources shows the presence of manifold bright-rimmed cloud (BRC) like structures 
with all their \textquoteleft heads\textquoteright\, 
pointing in almost the same direction - towards the center - suggestive of the 
fact that they could have been carved out by the radiation of central high-mass 
star(s); though alternatively, these illuminated bright-rims might just be the 
borders of normal density filaments carved out by the ionizing radiation and 
hence pointing towards the central region. There are very few YSOs associated 
with the heads of these BRCs, unlike what has been observed in many regions 
\citep{ogu10,chau09}.

\begin{figure}
\includegraphics[trim={1cm 3cm 0cm 5.5cm}, clip, scale=0.47]{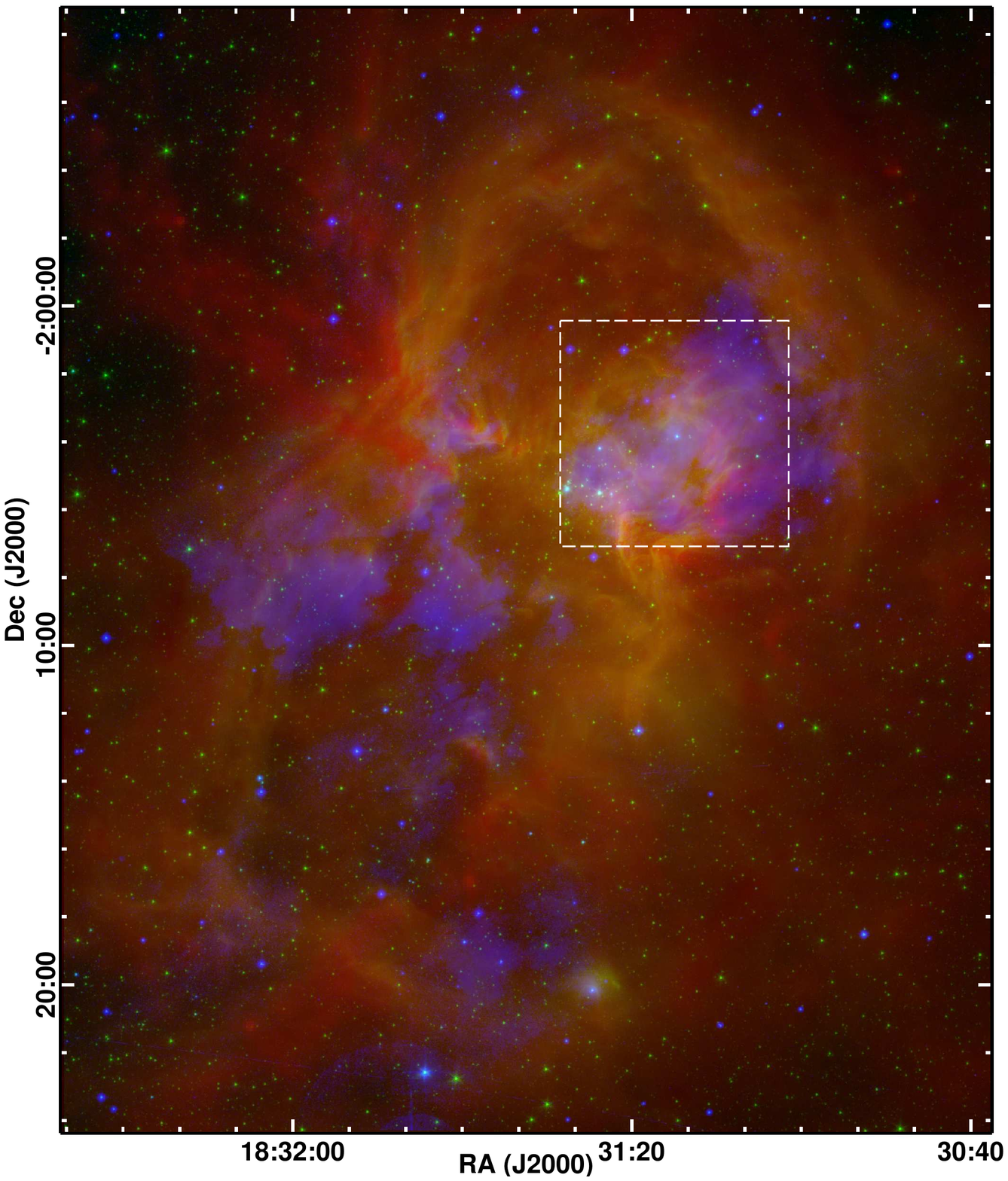} 
\caption{Colour composite image made using PACS 160 $\mu$m (red), IRAC 3.6 $\mu$m 
(green), and H$\alpha$ image (blue). Dashed white box shows the IRS\,5 nebular region.} 
\label{fig_WithHalpha}
\end{figure}

\subsubsection{Comparison with H$\alpha$ emission}
\label{section_SpatialDistributionHalpha} 

Figure \ref{fig_WithHalpha} shows a colour composite image made using PACS 160 $\mu$m 
(red), IRAC 3.6 $\mu$m (green), and H$\alpha$ image (blue) from SuperCOSMOS H$\alpha$ 
survey \citep{parker05}. The dashed white box in the image marks the IRS\,5 arc-shaped 
nebular region. 
The image shows the current star formation activity (traced by H$\alpha$ emission) 
vis-a-vis the dispersed molecular material by the high-mass sources (traced by PAH 
in 3.6 $\mu$m band). As can clearly be seen, there is no correlation between 
H$\alpha$ emission and the 3.6 $\mu$m emission; though this could partially be 
the effect of variable extinction. 
The H$\alpha$ emission regions are also not correlated with the location of our 
detected YSOs. H$\alpha$ is mainly present in the IRS\,5 nebular region, and in 
selected portions to the south of the midriff. Some of the H$\alpha$ emission is 
correlated with BRCs and the so-called elephant trunks (to the east). It is likely 
that the stellar objects in these H$\alpha$ emission regions are in extremely 
young stages.

\subsection{Estimating masses} 
\label{section_MassEstimation}  
We determined the mass of the cloud within the central cluster radius and the 
IRS\,5 region using the extinction map, assuming the empirically determined 
gas-to-dust ratio of $\langle N(H_{2})/A_V \rangle = 0.94 \times 10^{21}$ 
molecules cm$^{-2}$ mag$^{-1}$ \citep{ciardi98,lilley55,jenkins74,bohlin78}. 
This expression has been derived assuming that the total-to-selective 
extinction ratio (R$_V$) is 3.1, and that most of the gas is in form of 
molecular hydrogen \citep{frerking82, krumholz11}. 
For the central cluster region (which has a radius of 3\arcmin\, centered on 
IRS\,1A), first the column density at each pixel (of size 20\arcsec\,) was 
determined using the A$_V$ value in that pixel. Thereafter, column density 
values in all the pixels within the cluster area were integrated to obtain the 
total mass in that area. Same method was used to calculate the mass of the 
IRS\,5 region - with the area taken as that extending from the edge of the 
central cluster circumference till approximately the edge of the arc-shaped 
nebula. The areas used for the calculation of masses in the central cluster 
and the IRS\,5 region are mutually exclusive. The resulting masses obtained 
for the central cluster region and IRS\,5 region were $\sim$ 126 M$_{\odot}$ 
and $\sim$ 71 M$_{\odot}$, respectively. These estimates are most likely lower 
limits as a not unsignificant fraction of the molecular gas might have been 
expelled during the course of evolution of the cluster. Also, some of the gas 
will be present in form of the ionised gas. Another caveat to be kept in mind 
is that R$_V$ can, in general, have wide variations \citep{mathis90}.

\begin{figure}
\includegraphics[trim={1.7cm 1cm 0cm 3cm}, clip, scale=0.482]{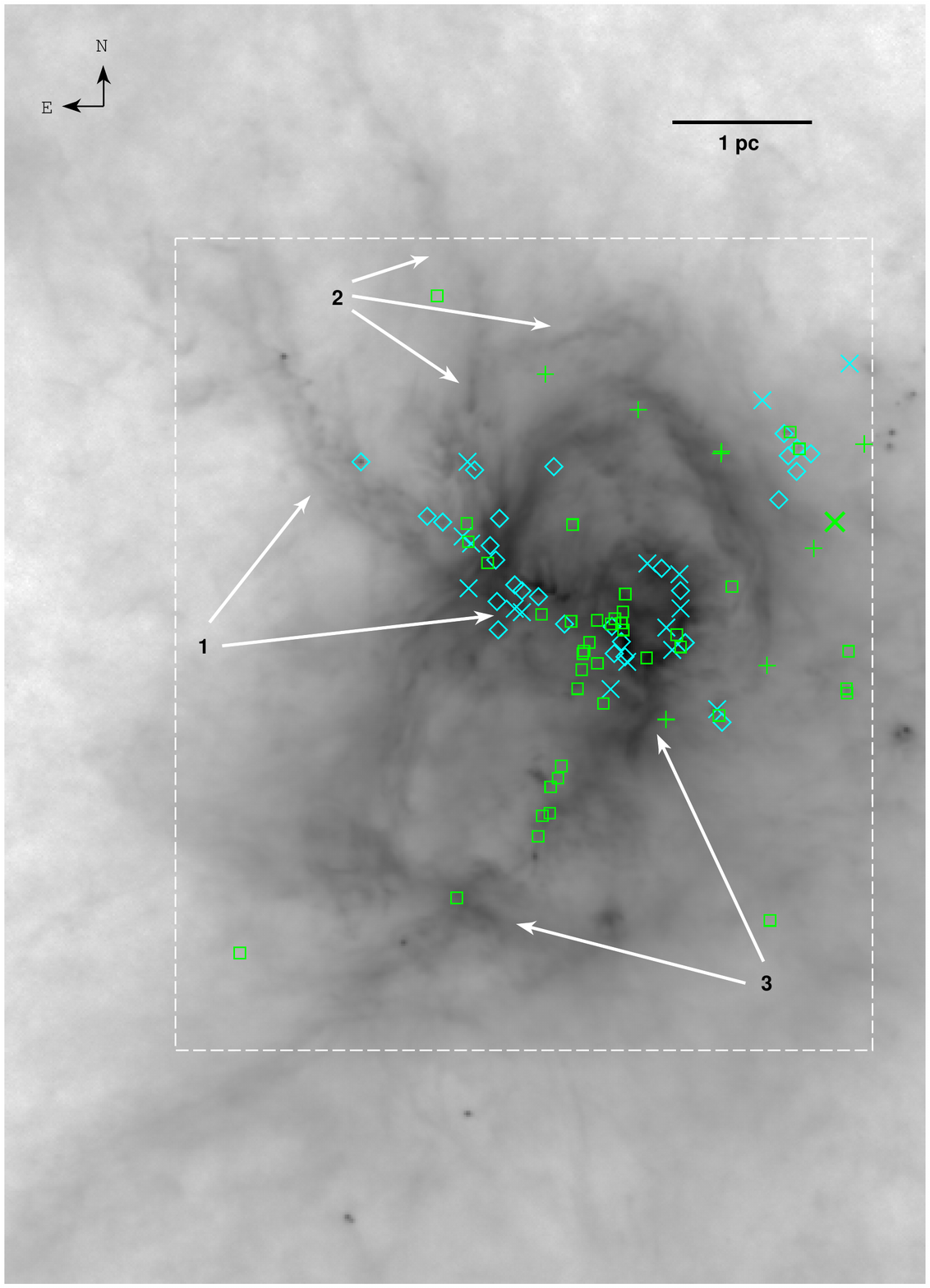} 
\caption{\textit{Herschel} PACS 160 $\mu$m image overlaid with the youngest sources 
in the region. Cyan diamond and cyan cross symbols denote protostars and starless 
cores, respectively, from \citet{mau11}. Green square symbols denote Class I sources.
Prominent filaments have been marked with arrows and labelled. Green plus symbols 
mark the radio sources from \citet{pir12}. Green cross marks the location of the 
molecular clump TGU 279-P7 from \citet{dob05}. White dashed box marks the area of 
our analysis in the paper.} 
\label{fig_FilamentaryStructure}
\end{figure}

\section{Morphology of the Region} 
\label{section_Morphology}

\subsection{Filamentary Structures} 
\label{section_FilamentaryStructures} 
Figure \ref{fig_FilamentaryStructure} shows \textit{Herschel} PACS 160 $\mu$m 
image of the larger region ($\sim$ 45.5\arcmin $\times$ 63.0\arcmin) with 
overlaid YSOs, and radio sources (in green plus symbols) from \citet{pir12}. The 
location of the molecular clump TGU\,279-P7 from \citet{dob05} has been marked in 
a green cross. Apart from the circular lobes which join at the midriff, filamentary 
structures are also seen emanating towards north-eastern and southern directions 
from the central region. In Figure \ref{fig_FilamentaryStructure}, three prominent
parsec-scale filaments, marked on the image, can clearly be made out. Due to high 
extinction in the region and by virtue of their intrinsically dense nature, the 
filamentary structures start becoming visible in emission only at \textit{Herschel} 
PACS 100 and 160 $\mu$m bands and at submillimeter SPIRE wavelengths 
\citep[see][for the image of the larger region encompassing W40]{kon10}. They can 
also be seen in absorption against the dominant PAH (plus the continuum) emission 
at 11.3 $\mu$m and 12.7 $\mu$m in WISE 12 $\mu$m band image (not shown here).

\begin{figure*}
\includegraphics[trim={3.7cm 9cm 1.5cm 11cm}, clip, scale=1.3]{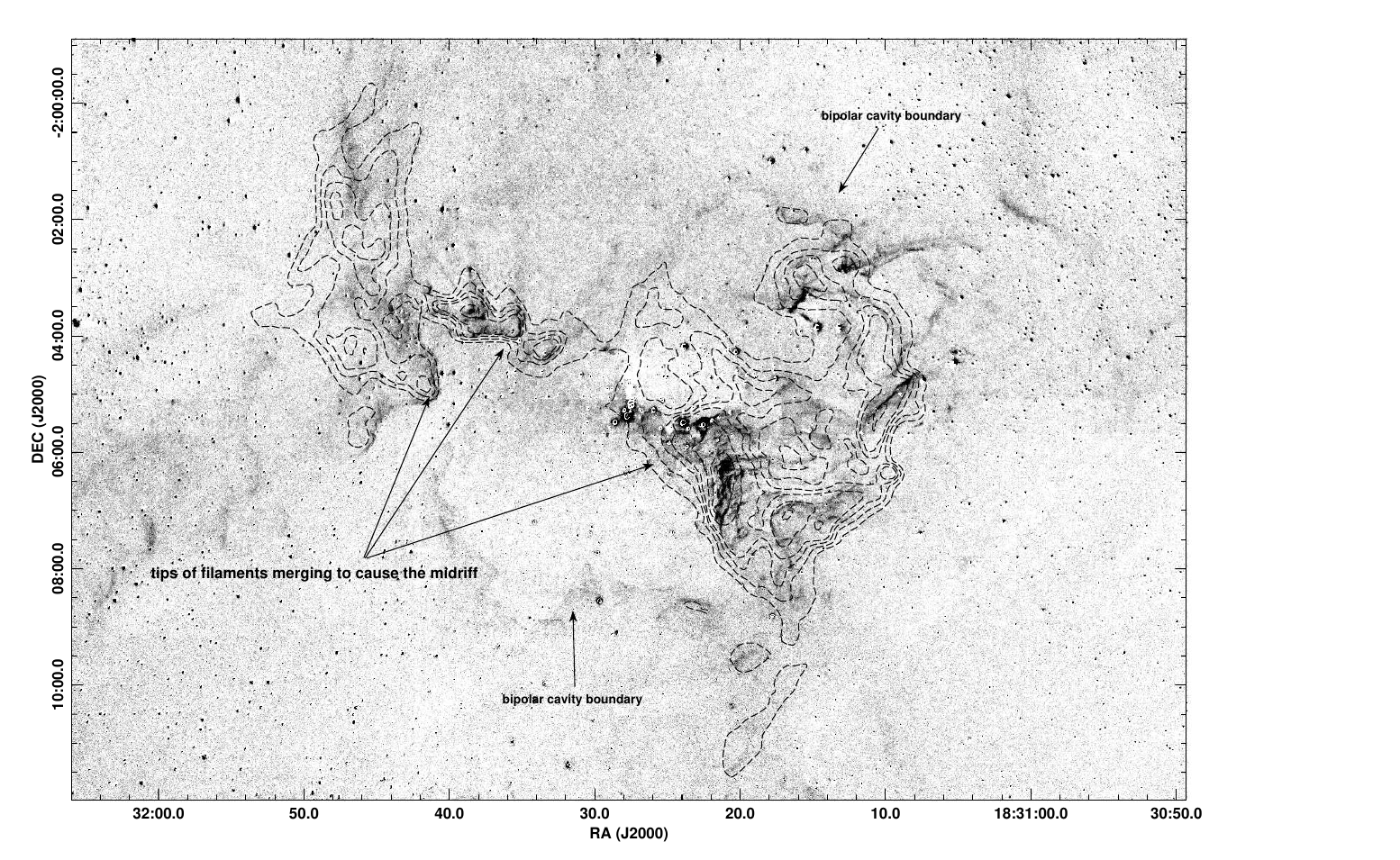} 
\figcaption{Continuum-subtracted 2.12 $\mu$m H$_2$ narrow-band image of the region 
in inverted grey-scale, with overlaid PACS 160 $\mu$m contours. The prominent features 
have been marked.} 
\label{fig_H2NarrowBand} 
\end{figure*}

Filament 1 is about $\gtrsim$ 2.4 pc long, and seems to be splitting into two parts 
near the midriff where it ploughs into the cloud. Possible signatures are the two 
bow shocks (the southern portion of filament 1, marked with an arrow) visible here
(also see Figure \ref{fig_SurfaceDensity}\textit{(left)} for clarity). 
Filament 2, which is comparatively diffuse, extends upto $\gtrsim$ 1.2 pc to the 
north of the central region, where it bifurcates into a filament to the north 
($\gtrsim$ 1.6 pc in length) and the other to the west ($\gtrsim$ 1 pc in length). 
Filament 3 is the longest visible structure, extending for $\gtrsim$ 3 pc from 
the western part of the central region, and seemingly cutting across the southern 
lobe of the bipolar nebula. 
According to the column density map of \citet{kon10}, the value throughout - for 
all the three filamentary structures - is of the order of 
$N_{H_{2}} \sim$ 10$^{22}$ cm$^{-2}$. The filamentary structures are also seen in 
dust temperature as well as extinction maps \citep[see their Figures 1 and 3]{bon10}. 

Figure \ref{fig_FilamentaryStructure} also shows overlaid YSOs, with Class I sources, 
protostars, and starless cores marked in green squares, cyan diamond symbols, and 
cyan crosses, respectively. As can be seen on the image, most of the sources seem 
to be distributed along two radially outward filaments 1 and 3 from the central 
region which has a concentration of these YSOs. Filament 2 hardly contains any of 
these youngest sources, probably because of its diffuse nature, suggesting that 
it is most likely still not Jeans critical. 

\citet{mye09} had proposed a hub-filament model according to which a central 
\textquoteleft hub\textquoteright\, with a column density of $\sim$ 10$^{22}$ cm$^{-2}$ 
radiates filaments which can be seen upto a column density of $\sim$ 10$^{21}$ cm$^{-2}$. 
It was suggested that the hub should have a minimum surface density of 25 YSOs pc$^{-2}$. 
The filamentary structures of the W40 region resemble this hub-filament structure. 
A possible \textquoteleft hub\textquoteright\, region is marked in 
Figure \ref{fig_CentralArea} by a white rectangle. This 
\textquoteleft hub\textquoteright\, region has a stellar density of 
$\sim$ 358 YSOs pc$^{-2}$ (total number of YSOs divided by total area), while the 
mean column density over this hub region is 
\textit{N}$_{H_{2}}$ $\sim$ 1.8 $\times$ 10$^{22}$ cm$^{-2}$ \citep{mau11}. Among 
the sources from our analyses and literature, a total of 12 Class I sources, 
6 protostars, and 126 Class II/III sources are located in this 
\textquoteleft hub\textquoteright\,. This translates to the fraction of youngest 
sources (Class I/protostars) being $\sim$ 0.125, and thus, according to the criteria 
used by \citet{mye09} (that this fraction should be $\geq$ 0.1), this region can 
be deemed \textquotedblleft young\textquotedblright.

\begin{figure*}
\centering
\includegraphics[trim={1cm 7cm 1cm 9cm}, clip, scale=0.90]{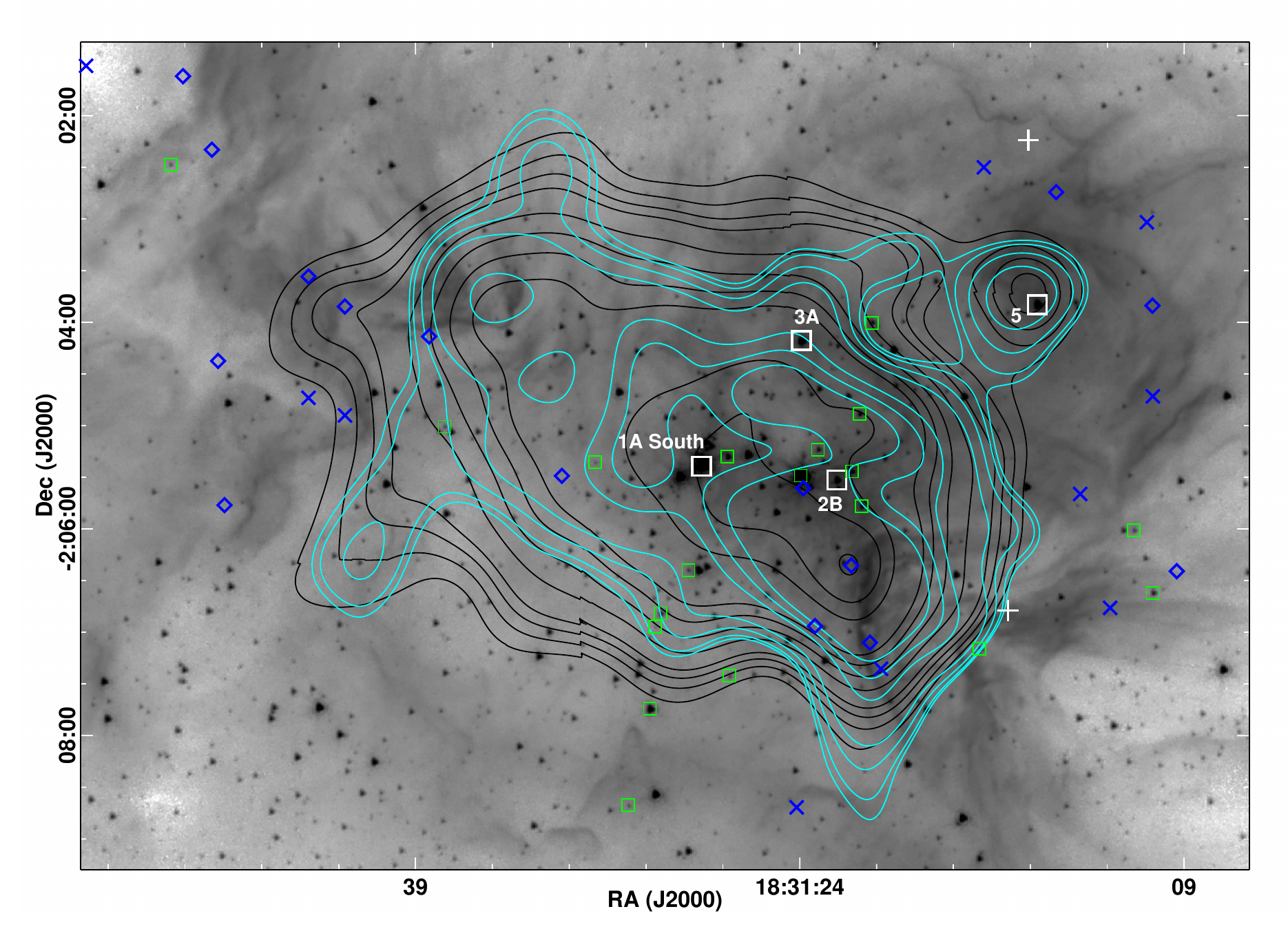}
\caption{\textit{Spitzer} 3.6 $\mu$m image overlaid with early type sources, YSOs 
and radio contours. 1280 MHz and 610 MHz contours have been marked in black and 
cyan, respectively. White squares with respective labels denote the early type 
stars in the region from \citet{shu12}. Plus symbols mark the $^{12}$CO(J=1-0) 
peaks \citep{smi85}. Blue diamond and blue cross symbols denote protostars and 
starless cores, respectively, from \citet{mau11}. Green square symbols denote 
Class I sources. 
1280 MHz contours are drawn at 4, 6, 8, 10, 15, 20, 25, 30, 50, 60, and 70 $\sigma$ 
($\sim$ 6.34 mJy/beam), while 610 MHz contours have been drawn at 4, 5, 7, 10, 20, 
25, and 30 $\sigma$ ($\sim$ 5.52 mJy/beam).} 
\label{fig_IRAC1withRadio} 
\end{figure*}

There are filaments to the west of this bipolar nebula which are a part of Serpens 
South cluster (not shown here). Though they appear to be spatially correlated with 
\object{W40} region, this could be a projection effect; as the distance to Serpens 
South is about 260 pc \citep{nak11}, while the distance to \object{W40} - though 
with varying values based on different techniques - has been estimated to be about 
500 pc \citep{shu12}.

\subsection{H$_2$ narrow-band}
\label{section_H2NarrowBandResults}
In Figure 9 
we display the continuum-subtracted H$_2$ narrow-band 
image using inverted grey scale. The brightest emission from the 
\textit{Herschel} PACS 160 $\mu$m image is overlayed using dashed 
contours. Several important features are marked on this figure which 
are explained below. Comparing H$_2$ emission with 160 $\mu$m contours
indicates that most of the observed H$_2$ emission traces the
boundaries of filaments and dense gas traced by the 160 $\mu$m
contours. The H$_2$ emission also reveals a faint bipolar shaped cavity
centered on the W40 embedded cluster. The northern lobe of this cavity
is clearly brighter than the southern lobe, however, the 160 $\mu$m
emission is also stronger in this northern region. BRCs can be seen 
along the southern boundary of this bipolar shaped cavity in 
IRAC images (see Figure \ref{fig_CentralArea}). The morphology of
the observed H$_2$ emission features are indicative of fluorescent
excited emission in most of the observed field. 
We note that the 
arc-shaped bipolar cavity in the northern region, partially surrounding
the IRS\,5 source, is also the region where several starless cores were
mapped by \citet{mau11} (see Section \ref{section_SpatialDistribution}). 
It may therefore represent a dense shell swept 
by the radiation from the central embedded cluster. While the lack of
collimated bipolar features indicate absence of shocked emission in
general, confirmation of the fluorescent or shocked nature for the
diffuse H$_2$ emission is not possible without obtaining spectroscopic
observations.

\subsection{Radio Morphology}
\label{section_RadioMorphology} 
Figure \ref{fig_IRAC1withRadio} shows the \textit{Spitzer} 3.6 $\mu$m image of the 
central portion of the W40 region. Overlaid are the 610 MHz and 1280 MHz radio 
contours from GMRT. The early-type stars in the region - IRS 1A South, 2B, 3A, 
and 5 have been marked with white squares and labelled. The $^{12}$CO(J=1-0) peaks 
from \citet{smi85} have been marked with plus symbols. It can clearly be seen that 
one of the $^{12}$CO(J=1-0) peaks lies at the western edge of the H~{\sc ii} region. 
The $^{13}$CO(J=2-1) and $^{13}$CO(J=3-2) maps from \citet{zhu06} also reveal the 
presence of molecular cloud to the west of ionizing region. This molecular cloud 
to the west of the H~{\sc ii} region has also been observed in various molecular 
lines and in continuum at millimeter wavelengths \citep{pirogov13}. The region 
thus appears to be density bounded to the west as is evidently borne out by the 
CO isotopologues' peaks. But, there is no such constraint towards the east, leading 
to the proliferation of ionizing photons towards this direction. This is indicative 
of blister morphology, also referred to as champagne flow \citep{teno79, whit79}, 
in the region. The radio contours depicted on the image also support this blister 
morphology for the region, with the head of the ionized flow towards the west 
(density-bounded), and an extended emission towards the east. The peak of 
1280 MHz H~{\sc ii} emission is about 1.24$^{'}$ to the south-west of the peak at 
408 MHz from \citet{gos70}, but this could be due to their lower beamsize 
($\sim$ 3$^{'}$). As can be seen in Figure \ref{fig_IRAC1withRadio}, there are no 
\textit{Spitzer} Class I YSOs near this peak, only Class 0/I sources or starless 
cores from \citet{mau11} are seen. Also, one of the Class 0/I sources from 
\citet{mau11} coincides with the 1280 MHz radio peak. This is a probable indication 
of some high-mass star formation going on in this region, which is also supported 
by the outflow seen by \citet{zhu06}. The timescale of the outflow from \citet{zhu06} 
($\sim$ 6 $\times$ 10$^{4}$ yr) is of the order of the age estimate from \citet{mau11} 
($\sim$ 4-9 $\times$ 10$^{4}$ yr) of Class 0/I sources. However, it is plausible 
that the \textquotedblleft outflow\textquotedblright\, observed by \citet{zhu06} 
might simply be the material swept-up and pushed into the bipolar shaped nebula. 
The H~{\sc ii} emission does not extend all the way up to the edge of the bipolar 
lobes (which marks the edge of the photo-dissociation region) or trace them at GMRT 
sensitivity levels. The extent of the central H~{\sc ii} region is almost same as 
the cluster radius at half-peak surface density 
(see Section \ref{section_SpatialDistribution} and Figure \ref{fig_SurfaceDensity}), 
which is expected. The central ionizing region is limited to $\sim$ 2$^{'}$ 
($\sim$ 0.29 pc) on either side of the midriff of the bipolar lobes. Though isolated, 
non-thermal radio emission sources have been discovered by \citet{pir12} (see 
Figure \ref{fig_FilamentaryStructure}), it seems unlikely that they are related 
to the W40 region due to their non-thermal nature. Another feature of note is the 
sub-region around IRS 5 which shows distinct radio emission with a roughly circularly 
symmetric ionisation region around this source.

\subsection{Overall Structure}
\label{section_OverallStructure} 
The presence of these various morphological structures support the schematic for 
this region by \citet[see their Figure 2]{val87}. The high-mass stars formed are 
probably located at the edge of the parental giant molecular cloud towards the 
observer, and as these sources have ionized the natal medium, the whole H~{\sc ii} 
region has broken out of the parental cloud. The filaments seem to be emanating 
from the midriff of the W40 region. This suggests that the density was high along 
the equatorial axis, as compared to the polar directions, leading to a high density 
contrast. Thus the Lyman continuum radiation has escaped along the polar directions, 
flinging the ionised gas (which forms the bipolar nebula) partially towards the 
observer. This flung-off material seems to have encountered more resistance towards 
the north - as seems likely by the smaller diameter of the northern lobe, more 
extincted regions towards the north, and brighter 160 $\mu$m emission from the 
northern lobe (see Figures \ref{fig_FilamentaryStructure}). This probably indicates 
that the main body of the cloud lies to the north of the W40 midriff region. The 
initial orientation and location of the high-mass stars seem such that this presence 
of molecular cloud to the north has proven harder to disrupt, while the southern 
thin shell of molecular material (between the high-mass stars and the ISM) has been 
blown away relatively easily. We note that the southern lobe could be appearing 
larger than the northern lobe due to the projection effect from the viewing angle, 
i.e. the region might not be being viewed exactly face-on, but from a non-zero 
inclination angle.  
As has also been evidenced in \citet{beau10} for various other regions, these 
northern and southern lobes could be more like rings than spheres, opening up the 
possibility that the molecular cloud could be oblate or sheet-like. The morphology 
of the filamentary structure shows its complete extent till the filaments join the 
central region. Since the filaments can be seen right till they join the midriff, 
they are most likely in front of the progenitor molecular cloud.

\section{Physical Parameters}
\label{section_PhysicalParameters} 

\subsection{Overall Region} 
\label{section_OverallRegion} 
\begin{figure}
\centering
\subfigure
{
\includegraphics[trim={0cm 0cm 7cm 23cm}, clip, scale=0.82]{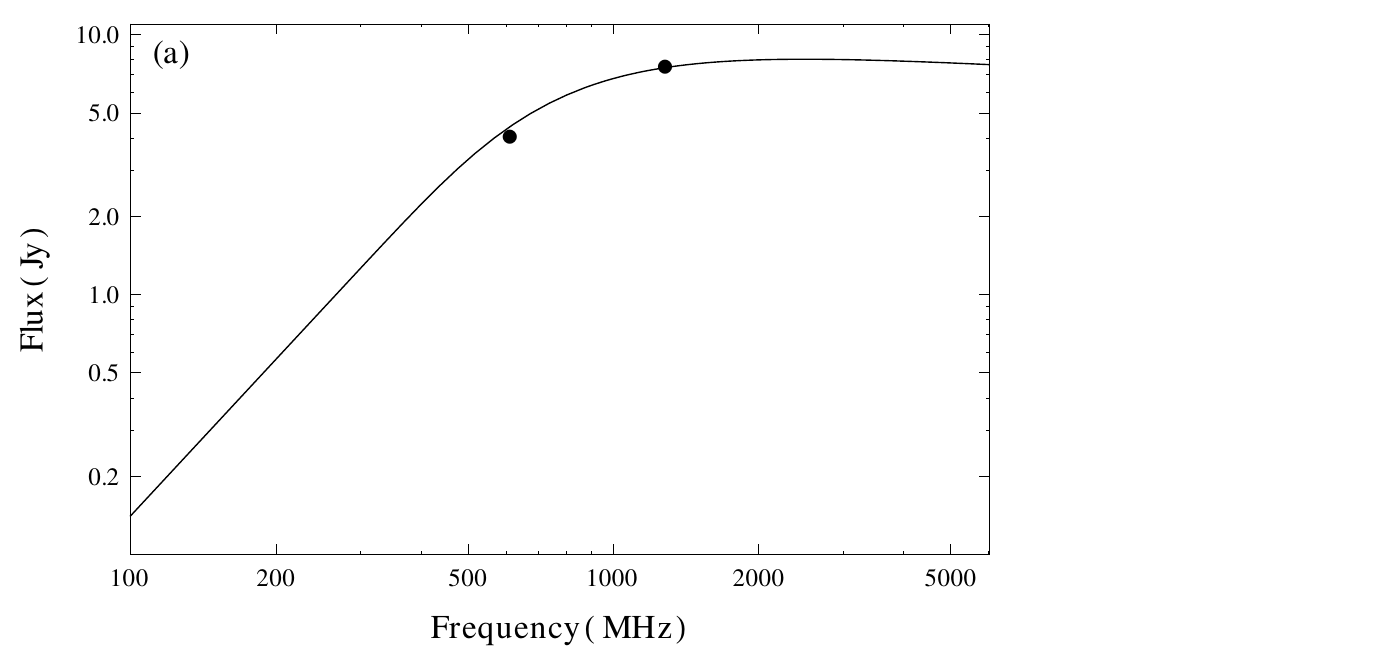}
\label{fig_EM-fit-all}
}
\subfigure
{
\includegraphics[trim={0cm 0cm 7cm 23cm}, clip, scale=0.82]{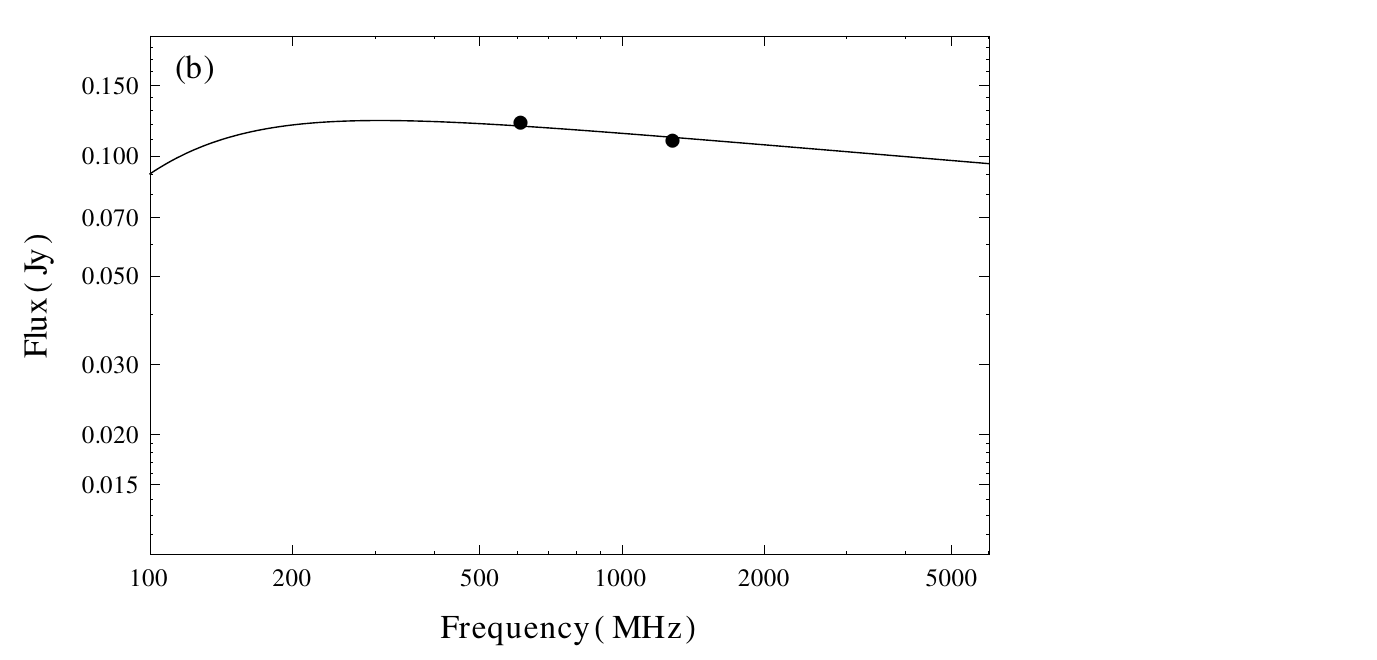}
\label{fig_EM-fit-IRS5}
}
\caption{The fitted flux density model (solid line) for (a) the entire region, 
and (b) the IRS 5 region. The data points at 610 MHz and 1280 MHz have been 
marked with black solid circles.}  
\label{fig_EM-fit}
\end{figure}

Radio continuum observation data was used to derive the characteristic physical 
parameters like emission measure, electron density, dynamical age, and 
str\"{o}mgren radius for the region. Since this region contains multiple radio 
sources \citep{rod08}, we carried out a global analysis using the low resolution 
images ($\sim$ 45$^{''} \times$ 45$^{''}$) to get an estimate of the parameters. 
Integrated flux density was calculated for both the frequencies within 4$\sigma$ 
contours using the AIPS task ``TVSTAT''; giving us a value of 4.07 Jy and 7.55 Jy 
at 610 MHz and 1280 MHz, respectively. Thereafter, using the model for free-free 
emission of \citet{mez67}, the flux density in a region is given by 
\citep[adapted from][]{mezEtAl67}: 

\begin{equation}
 S_{\nu}=3.07\times10^{-2}T_{e}\nu^{2}\Omega(1-e^{-\tau(\nu)})
\end{equation}

\begin{equation}
 \tau(\nu)=1.643\times10^{5}aT_{e}^{-1.35}\nu^{-2.1}n_{e}^{2}l
\end{equation}

where, $S_{\nu}$ is the integrated flux density in Jy, $T_{e}$ is the electron 
temperature of the ionized core in K, $\nu$ is the frequency in MHz, $n_{e}$ is 
the electron density in cm$^{-3}$, $l$ is the extent of the ionized region in pc, 
$\tau$ is the optical depth, $a$ is the correction factor, and $\Omega$ is the 
solid angle subtended by the beam in steradian here. $n_{e}^{2}l$, called the emission 
measure, measures the optical depth in the medium (in cm$^{-6}$\,pc). For our 
calculation, we took $T_{e}$ to be 8500 K \citep{sha70,qui06}, and thus the 
correction factor $a$ as 0.99 \citep[Table 6]{mez67}. Using the two data points 
for 610 MHz and 1280 MHz, we fit the above equations using a non-linear regression 
with emission measure ($n_{e}^{2}l$) as a free parameter 
(Figure \ref{fig_EM-fit-all}). This yielded the value of emission measure as 
1.63 $\pm$ 0.09 $\times$ 10$^{6}$ cm$^{-6}$\,pc. The extent ($l$) of the central 
H~{\sc ii} region is approximately 7.0$^{'}$, equivalent to about 1.02 ($\pm$ 0.41) 
pc at a distance of 500 ($\pm$ 200) pc; and thus the electron density turns out 
to be $\sim$ 1265 $\pm$ 218 cm$^{-3}$ for this central region. 

The Lyman continuum luminosity (photons s$^{-1}$) required for ionizing the gas 
was calculated by using the formulation of \citet[see their Equation 5]{mor83} 
for the optically thin regime. As is evident from Figure \ref{fig_EM-fit-all}, 
610 MHz and 1280 MHz emission lie in the optically thick and thin regimes, 
respectively. Therefore, using the flux density of 7.55 Jy estimated at 1280 MHz, 
the Lyman continuum luminosity was calculated to be $\sim$ 1.67 $\times$ 10$^{47}$ 
photons s$^{-1}$. \citet{shu12} have estimated that the radio emission is dominated
by an O9.5 type star (IRS 1A South, see Figure \ref{fig_IRAC1withRadio}) with a 
luminosity of 6.3 $\times$ 10$^{47}$ photons s$^{-1}$. They place an upper limit on 
the Lyman continuum luminosity to be 1.6 $\times$ 10$^{48}$ photons s$^{-1}$. 
\citet{smi85}, using IR and low resolution ($\sim$ 11$^{'}$) radio observation of 
\citet{alt70} inferred a luminosity of $\sim$ 1.5 $\times$ 10$^{48}$ photons s$^{-1}$. 
Our Lyman continuum luminosity estimate differs by a factor of the order of unity 
from the calculated flux of an O9.5 type star in \citet{shu12} and the tabulated 
value from \citet{martins05}. However, it is about 10\% of the value from \citet{smi85} 
and of the upper limit in \citet{shu12}. This could be because of absorption of photons 
by the dust in the region, implying the presence of dense gas in the central region. 
As has also been noticed by \citet{kurtz99}, presence of dust absorption can lead to 
underestimation of radio luminosity. 
\citet{arthur04} have estimated that dust absorption can range from 60\% upto 
96\% of the ionizing photons. Similar underestimation of ionizing flux by radio data 
has been observed for other regions in \citet{brand11, alvarez04}. 
The luminosity calculated here, therefore, is most likely just the lower limit.

\begin{figure}
\centering
\subfigure
{
\includegraphics[trim={0cm 0cm 7cm 23cm}, clip, scale=0.80]{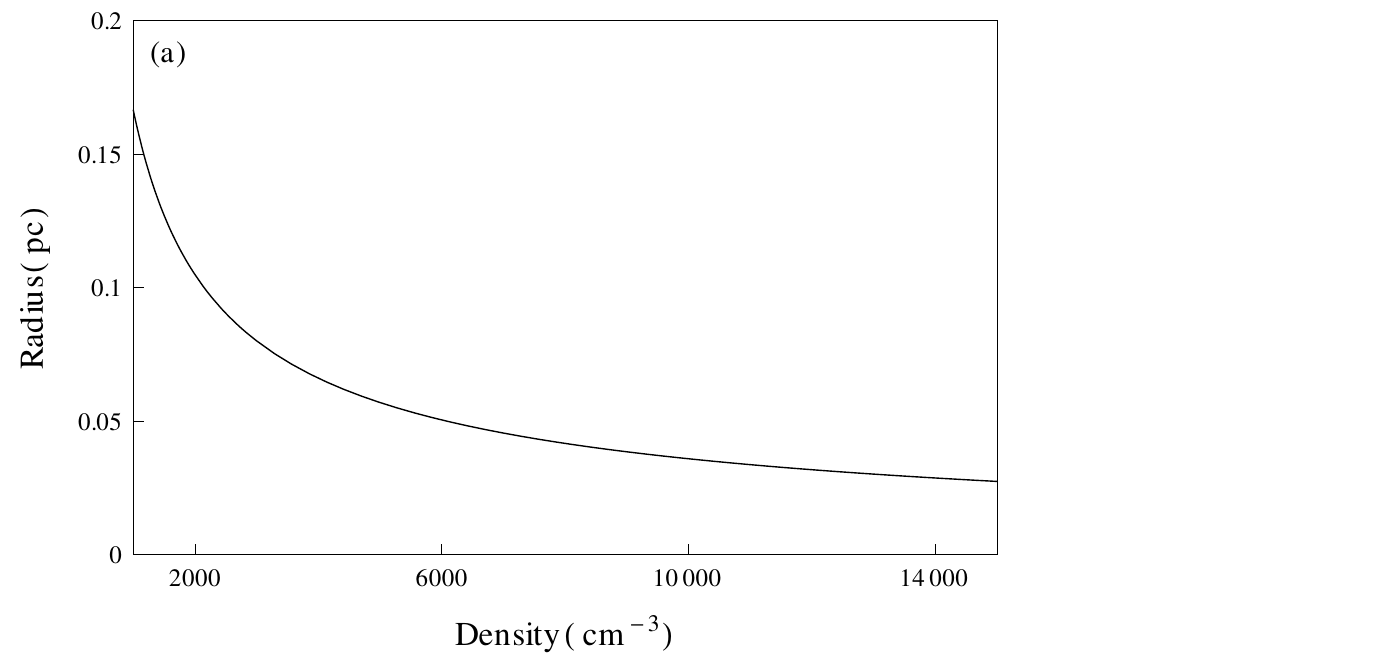}
\label{fig_Radius}
}
\subfigure
{
\includegraphics[trim={0cm 0cm 7cm 23cm}, clip, scale=0.80]{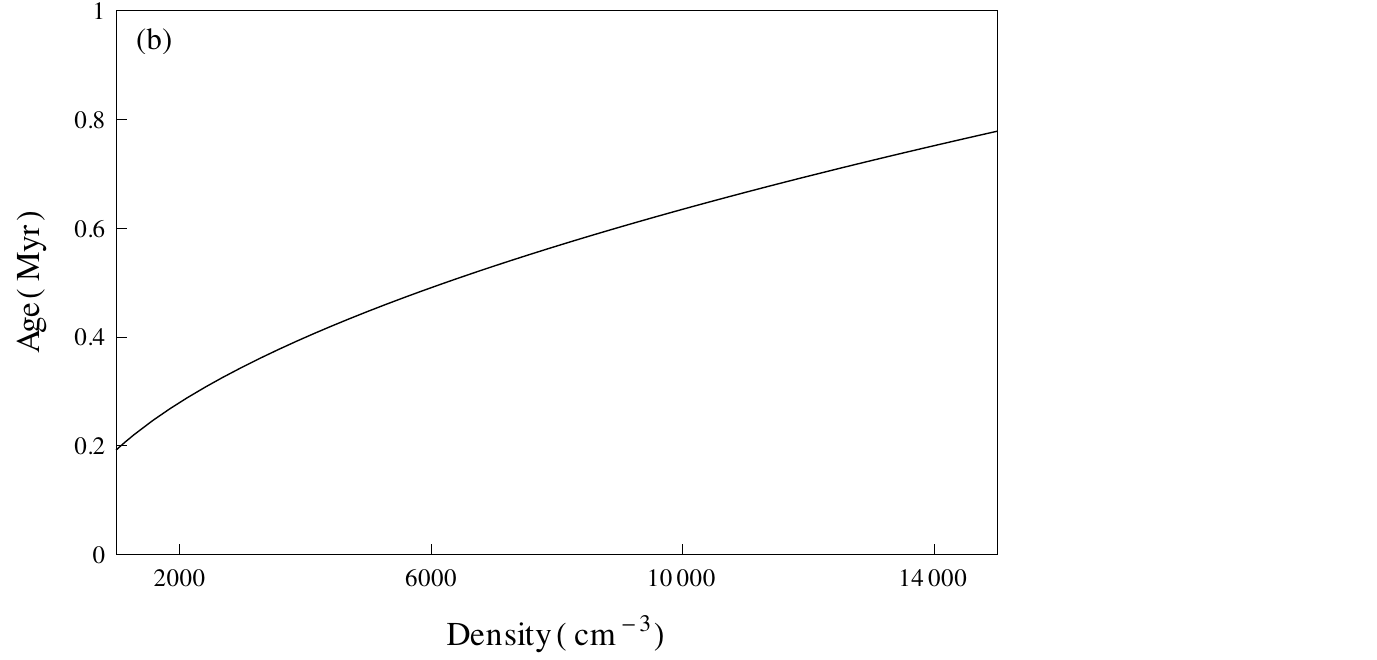}
\label{fig_Age} 
}
\caption{Graph showing the variation of (a) Str\"{o}mgren radius and (b) dynamical age with ambient density for the overall
photoionized region.} 
\label{fig_RadiusAge}
\end{figure}

In the first stage during the ionization of a region by massive star(s), the 
ionization front propagates outwards leading to an expansion of the H~{\sc ii} 
region, till an equilibrium is reached between the number of ionizations and 
recombinations. The radius of the H~{\sc ii} region at this point, assuming that 
the ambient medium has spatially uniform density, is given by \citep{str39} : 

\begin{equation}
 R_{s}=\left(\frac{3S_{Lyman}}{4\pi n_{o}^{2}\beta_{2}}\right)^{1/3}
\label{equation_StromgrenRadius} 
\end{equation}

where $R_{s}$ is called the Str\"{o}mgren radius (in cm). $n_{o}$ is the initial 
ambient density (in cm$^{-3}$), and $\beta_{2}$ is the total recombination coefficient 
to the first excited state of hydrogen. The value of $\beta_{2}$, for a $T_{e}$ 
of 8500 K, was taken to be 2.94$\times$10$^{-13}$ cm$^{3}$ s$^{-1}$ \citep{war11,sta05}. 
In the second stage of expansion, after the Str\"{o}mgren radius has been reached, 
the shock front overtakes the ionization front due to the vast pressure difference 
between material inside and outside of the ionization front. In this stage, the 
radius of the H~{\sc ii} region is given by \citep{spi78} : 

\begin{equation}
 R(t)=R_{s}\left(1+\frac{7c_{II}t}{4R_{s}}\right)^{4/7}
\label{equation_DynamicalAge}  
\end{equation} 

where $R(t)$ denotes the extent of the H~{\sc ii} region at time $t$, and $c_{\sc II}$ 
is the speed of sound, taken as 11$\times$10$^{5}$ cm s$^{-1}$ \citep{sta05}. Since 
we have no way of gauging the initial ambient density ($n_o$) of the molecular 
cloud, we have plotted the parameters for ambient density ranging from 1000 to 
15000 cm$^{-3}$ to get an estimate of the ranges. Figure \ref{fig_RadiusAge} shows 
Str\"{o}mgren radius ($R_{s}$) and dynamical age ($t$) plotted as a function of 
initial ambient density of the medium. Str\"{o}mgren radius varies from $\sim$ 
0.17 to 0.03 pc, while the dynamical age varies from $\sim$ 0.19 to 0.78 Myr. 
If we were to take the higher estimates of $S_{Lyman}$ (the values discussed above), 
then, for a fixed density, the $R_s$ will increase and the $t$ will decrease. 
It should be noted that there is more than one massive star in the region, and each 
massive star will have its own Str\"{o}mgren radius and age. The multiple 3.6 cm 
VLA sources found in the region \citep{rod10} will also contribute to the total 
flux. Also, the radio morphology is not spherical here. Therefore the above 
calculations should be taken as a representative value for the Str\"{o}mgren radius 
and age estimate, and $n_{e}$ should be taken as the averaged out electron density 
for the ionized region.

\subsection{IRS 5 Region} 
\label{section_IRS5_RadioCalculations} 
Since radio emission around IRS 5 source is distinct from the rest of the morphology 
(see Figure \ref{fig_IRAC1withRadio}), we can carry out a separate calculation to 
determine the physical parameters around this source. For this, we first determined 
the 610 MHz and 1280 MHz integrated flux density in this region in a circle of 
50\arcsec\ radius centered at the peaks of respective frequencies. This yielded 
a flux density of 121 and 109 mJy for 610 MHz and 1280 MHz, respectively. \citet{rod10} 
have detected IRS 5 source using VLA observations at 3.6 cm. However, they calculated 
a flux of 0.92 mJy for sub-arcsecond area, while here we are concerned with the 
extended emission. Another VLA source from \citet[Source 2 in their Table 2]{rod10} 
lies close to the IRS 5 source. 
However, it has a flux of 0.90 mJy (at 3.6 cm), and is hence insignificant for our 
calculation here. Following a similar set of steps as above 
(Section \ref{section_OverallRegion}), we determined the emission measure by fitting 
the free-free emission SED (see Figure \ref{fig_EM-fit-IRS5}) to be 
$\sim$ 2.01 $\pm$ 0.04 $\times$ 10$^{4}$ cm$^{-6}$ pc. The extent of radio contours 
for this region is $\sim$ 100\arcsec\ ($\sim$ 0.24 $\pm$ 0.10 pc at a distance of 
500 $\pm$ 200 pc) which gives the electron density as $\sim$ 288 $\pm$ 55 cm$^{-3}$. 

We estimated the Lyman continuum luminosity using the formulation of \citet{mor83} 
and 1280 MHz data point as in Section \ref{section_OverallRegion} (in this case, 
since 610 MHz data point is in the optically thin regime too, it yields similar 
value), which results in a value of $\sim$ 2.40 $\times$ 10$^{45}$ photons s$^{-1}$. 
Comparing log(Lyman continuum luminosity) (45.38) with the tabulated values from 
\citet[Table II]{pan73}, we get the most probable spectral type of IRS 5 as B1V, 
in agreement with the values from \citet{shu12}. The slightly lower value of the 
calculated Lyman continuum luminosity, as compared to the value from \citet{pan73} 
is probably due to the absorption of photons by dust in the region.  

Though the radio contours are seen within $\sim$ 50\arcsec\ radius of the respective 
radio peaks (Figure \ref{fig_IRAC1withRadio}), it can be seen that the ionised 
region extends beyond that. This weak extended emission region is traced by PAH 
emission at 7.7 $\mu$m and 8.7 $\mu$m in IRAC Ch4 and at 3.3 $\mu$m in IRAC Ch1 
(which, it must be kept in mind, also contain continuum emission) (also see 
Figure \ref{fig_CentralArea}). Most of Class 0/I sources and starless cores from 
\citet{mau11} are approximately along the circumference of a circle of radius 
$\sim$ 80\arcsec\ centered at respective radio peaks. Therefore, taking the extent 
to be 160\arcsec\, and the ambient density ($n_{o}$) to be a typical value of 
$\sim$ 1000 cm$^{-3}$ \citep{sta05}, we use equation \ref{equation_StromgrenRadius} 
in conjunction with equation \ref{equation_DynamicalAge} to obtain the dynamical 
age of the H~{\sc ii} region around the IRS\,5 arc-shaped nebula to be $\sim$ 
0.11 Myr. We note that though the ionised region appears more extended than what 
we have assumed in the above calculation, and the ambient density of the medium 
could be larger, a higher value of both these parameters will only increase the 
value of the dynamical age. Therefore, the 0.11 Myr should be considered a lower 
limit for the dynamical age of this sub-region.

\section{Star Formation Scenario} 
\label{section_Star-formation-scenario} 

The nature of the YSOs with their respectively distinct spatial distributions 
suggest a scenario in which (spontaneous) star formation has taken place at 
different epochs for YSOs at different evolutionary stages. It is very likely 
that Class II/III and the central early-type sources formed in an earlier epoch 
as the central region became Jeans critical, 
followed by the formation of younger sources - Class 0/I sources and starless 
cores from \citet{mau11} and Class I YSOs from our analysis - along filaments 
in a later epoch. The lifetimes of the dominant early type star in this region 
\citep[O9.5; ][]{shu12} is of the order of $\sim$ 10 Myr \citep{sta05}, while the 
lifetimes of Class II/III sources is also of the order of a few Myr. The fact that 
the two lifetimes are comparable and that this region is not very old, makes it 
likely that the majority of the Class II/III and central high-mass sources formed 
at the same epoch. \citet{rod08}, based on the velocity of molecular hydrogen, have 
opined that star formation might have been initiated several Myrs ago due to an 
external shock. It is probable that this could have played a role too and affected 
the earlier epoch of star formation - one which led to the formation of Class II/III 
sources and central high-mass stars. This seems to have been followed by star 
formation leading to younger YSOs. Alternatively, it is possible that the formation 
of high-mass stars in the center could have driven a compressional wave into the 
individual filaments, leading the high-density regions in the filaments to become 
Jeans critical, thus subsequently initiating the formation of Class I sources, 
protostars, and starless cores. 

The stellar distribution seen has also been observed by \citet{sch12}, who, using 
the Rosette molecular cloud as a template, have done a density analysis to conclude 
that almost all IR clusters lie at the junction of filaments. Simulations by  
\citet{dale12} also show that the star formation during the course of evolution 
of molecular clouds to be mostly confined to the filaments and their junctions. 

\citet{bon10}, assuming the distance to W40 to be 260 pc, have estimated the 
mass of the molecular cloud (G28.74+3.52) associated with the W40 H~{\sc ii} 
region to be about 1.1 $\times$ 10$^4$ M$_{\odot}$ using 2MASS extinction map. 
The mass estimated using extinction map depends on the distance as a square of 
it. \citet{shu12} have recently ruled out distances below 340 pc for this region. 
Also, mass estimation in high optical depth regions can often get underestimated 
using this method. So, the mass estimated by \citet{bon10} is most likely a 
lower limit and the actual mass could be slightly higher, probably between a 
few 10$^4$ M$_{\odot}-$10$^5$ M$_{\odot}$. 

The IRS\,5 region (see Figure \ref{fig_CentralArea}) also seems to be hosting  
fresh star formation at the edge of the cavity excavated by the high-mass source. 
The starless cores and Class 0/I sources from \citet{mau11} are found to be 
distributed along the edge of the H~{\sc ii} region of this arc-shaped nebula. The  
arc of (seemingly) collected molecular material, along the edge of this arc-shaped 
nebula, has also been observed in various molecular lines \citep{pirogov13}, 
as well as at submillimeter wavelengths \citet{mau11} and H$_2$ narrow-band image
(see Section \ref{section_H2NarrowBandResults}). The farthest YSO from the 
IRS\,5 source in Figure \ref{fig_CentralArea} is located at a distance of $\sim$ 
1.8\arcmin\ ($\sim$ 0.26 pc). Therefore, using the isothermal sound speed of 
11\,km\,s$^{-1}$ for the ionised region (see Section \ref{section_OverallRegion}), 
it can be seen that the high-mass star could have lead to the sweeping-up of 
material which subsequently might have provided impetus to the formation of 
protostars and starless cores in $\sim$ 0.02 Myr. From the radio analysis 
(see Section \ref{section_IRS5_RadioCalculations}), the lower limit for the dynamical 
age of this sub-region was calculated to be $\sim$ 0.11 Myr. The ages of Class\,0 and 
Class\,I YSOs are generally of the order of a few 0.01 Myr and 0.1 Myr, respectively, 
and as such, their formation (and those of starless cores) could have been influenced 
by the expanding H~{\sc ii} region. We add a caveat that though the numbers are 
consistent with a scenario where the high-mass star could have influenced the star 
formation at the periphery, this might not be the correct scenario necessarily owing 
to possible complex and hitherto unexamined morphological structures in this sub-region. 

Based on the above discussions, the following appears to be a plausible chronology of 
star formation in this region. As the filaments formed and joined, they led to the 
formation of dense core at their junction, referred to as midriff in this paper. 
This seems to have initiated the formation of the central high-mass star(s) and other 
Class II/III sources. At this point, an external shock - as mentioned by \citep{rod08} - 
might have given a boost to the star formation going on. The high-mass source IRS 1A
South was formed at the junction, as well as the entire cluster which is centered upon 
this source. Due to lower density along the polar direction as opposed to the equatorial 
direction, when the high-mass star(s) formed, the Lyman continuum radiation escaped 
along the polar directions leading to the formation of the bipolar nebula. Fresh 
(spontaneous) star formation seems to be going on along the filamentary structures and 
their junction (albeit at lower stellar density) as evidenced by the younger sources in 
Class 0/I stages and starless cores. A few of BRCs and elephant-trunks exhibiting 
H$\alpha$ emission to the south of the midriff might have undergone evolution under the 
influence of the ionizing region which has chiselled them out. The high-mass source 
IRS\,5, which is distinct from the rest of the radio emission region, seems to have led 
to the formation of a separate arc-shaped nebula, and swept-up material to its periphery. 

Future observations and analysis of the whole region in various molecular lines, radial 
velocity measurements, analysis of cluster properties using minimum-spanning tree method, 
understanding the mass and luminosity functions of the region, observations of indivdual 
BRCs, and spectral observations of other NIR bright sources will help in putting the 
star formation scenario on a firm footing.

\section{Conclusions} 
\label{section_Conclusions} 
In this paper, we carried out a multiwavelength study of the Galactic star-forming 
W40 H~{\sc ii} region. Our main conclusions are as follows : 
\begin{enumerate}
\item 
Using the MIR data from \textit{Spitzer} in conjunction with NIR data from UKIRT, 
1202 YSOs were identified in the region, out of which 40 are Class I sources and 
1162 are Class II/III sources. Analysis of the YSO distribution and nearest-neighbour 
surface density yields the cluster radius as $\sim$ 0.44 pc and peak surface density 
as 650 pc$^{-2}$. Mass calculation using 
extinction map yields a value of $\sim$ 126 M$_{\odot}$ within this radius. 
\item 
The filamentary structures were examined to reveal 3 parsec scale filaments 
emanating from the midriff. Two of them (filaments \textquoteleft 1\textquoteright\, 
and \textquoteleft 3\textquoteright\, in our labelling) contain most of the youngest 
YSOs aligned along their lengths. Filament \textquoteleft 2\textquoteright\, was 
found to be relatively diffuse with hardly any of the youngest YSOs. 
\item 
SED fitting using the radio continuum emission at 610 MHz and 1280 MHz for the total 
emission region yielded the value of electron density to be $\sim$ 
1265 $\pm$ 218 cm$^{-3}$ and the total Lyman continuum luminosity to be $\sim$ 
1.67 $\times$ 10$^{47}$ photons s$^{-1}$. The dynamical age, for the ambient density 
ranging from 1000 to 15000 cm$^{-3}$, ranges from $\sim$ 0.19 to 0.78 Myr. 
\item 
The IRS\,5 arc-shaped nebular region was found to be distinct from the rest of the 
emission region, and was thus examined in radio separately. The electron density 
was obtained to be 288 $\pm$ 55 cm$^{-3}$ and a lower limit on the dynamical age 
to be $\sim$ 0.11 Myr. A comparison of radio continuum photon luminosity with the 
tabulated values from \citet{pan73} shows IRS\,5 to be of B1V spectral type, reaffirming 
previous estimate from \citet{shu12}. Extinction map gives a value of 
$\sim$ 71 M$_{\odot}$ as the mass for this arc-shaped nebula. 
\item 
The star formation seems to have taken place in two successive epochs, with the 
formation of relatively-older Class II/III and the central high-mass sources - 
resulting in a cluster which is centered around the high-mass star IRS 1A South 
- followed by that of the youngest sources (Class 0/I, and starless cores). A 
distinct case is of the IRS\,5 nebular region, where material seems to have been 
swept-up to the edge of this arc-shaped nebula by the expanding H~{\sc ii} region 
of the IRS\,5 source. 
\end{enumerate}

We thank the anonymous referee for a thorough and critical reading of the manuscript, 
and for the suggestions which helped in improving this paper. 
K.K.M., M.S.N.K., D.K.O., and M.R.S. acknowledge support from Marie Curie IRSES 
grant (230843) under the auspices of which this work was carried out. K.K.M. would 
like to thank Hendrik Linz, MPIA (Heidelberg) for his invaluable help in obtaining 
the \textit{Herschel} data. This research made use of data products from the 
\textit{Spitzer} Space Telescope Archive. These data products are provided by the 
services of the Infrared Science Archive operated by the Infrared Processing and 
Analysis Centre/California Institute of Technology, funded by the National Aeronautics 
and Space Administration and the National Science Foundation.


\begin{thebibliography}

\bibitem[Allen et al.(2006)]{all06} Allen, L., Bourke, T., Brooke, T., et al.\ 2006, Spitzer Proposal, 30574 

\bibitem[Altenhoff et al.(1970)]{alt70} Altenhoff, W.~J., Downes, D., Goad, L., Maxwell, A., \&
                                        Rinehart, R.\ 1970, \aaps, 1, 319 

\bibitem[Alvarez et al.(2004)]{alvarez04} Alvarez, C., Feldt, M., Henning, T., et al.\ 2004, \apjs, 155, 123 

\bibitem[Arthur et al.(2004)]{arthur04} Arthur, S.~J., Kurtz, S.~E., Franco, J., \& Albarr{\'a}n, M.~Y.\ 2004, 
                                        \apj, 608, 282 

\bibitem[Beaumont \& Williams(2010)]{beau10} Beaumont, C.~N., \& Williams, J.~P.\ 2010, \apj, 709, 791 
                                        
\bibitem[Bertin \& Arnouts(1996)]{bert96} Bertin, E., \& Arnouts, S.\ 1996, \aaps, 117, 393 
                                        
\bibitem[Bessell \& Brett(1988)]{bes88} Bessell, M.~S., \& Brett, J.~M.\ 1988, \pasp, 100, 1134 

\bibitem[Bohlin et al.(1978)]{bohlin78} Bohlin, R.~C., Savage, B.~D., \& Drake, J.~F.\ 1978, \apj, 224, 132 

\bibitem[Bontemps et al.(2010)]{bon10} Bontemps, S., Andr{\'e}, P., K{\"o}nyves, V., et al.\ 2010, \aap, 518, L85 

\bibitem[Brand et al.(2011)]{brand11} Brand, J., Massi, F., Zavagno, A., Deharveng, L., \& Lefloch, B.\ 2011, 
                                      \aap, 527, A62  

\bibitem[Cantalupo et al.(2010)]{can10} Cantalupo, C.~M., Borrill, J.~D., Jaffe, A.~H., Kisner, T.~S., 
                                        \& Stompor, R.\ 2010, \apjs, 187, 212  

\bibitem[Casali et al.(2007)]{cas07} Casali, M., Adamson, A., Alves de Oliveira, C., et al.\ 2007, \aap, 467, 777 

\bibitem[Casertano \& Hut(1985)]{casertano85} Casertano, S., \& Hut, P.\ 1985, \apj, 298, 80 

\bibitem[Chauhan et al.(2009)]{chau09} Chauhan, N., Pandey, A.~K., Ogura, K., et al.\ 2009, \mnras, 396, 964 

\bibitem[Chavarr{\'{\i}}a et al.(2008)]{chavarria08} Chavarr{\'{\i}}a, L.~A., Allen, L.~E., Hora, J.~L., 
                                                     Brunt, C.~M., \& Fazio, G.~G.\ 2008, \apj, 682, 445  

\bibitem[Ciardi et al.(1998)]{ciardi98} Ciardi, D.~R., Woodward, C.~E., Clemens, D.~P., Harker, D.~E.,
                                        \& Rudy, R.~J.\ 1998, \aj, 116, 349 

\bibitem[Cohen et al.(1981)]{coh81} Cohen, J.~G., Persson, S.~E., Elias, J.~H., \& Frogel, J.~A.\ 1981, \apj, 249, 481 

\bibitem[Dale et al.(2012)]{dale12} Dale, J.~E., Ercolano, B., \& Bonnell, I.~A.\ 2012, \mnras, 424, 377 

\bibitem[Davis et al.(2007)]{davis07} Davis, C.~J., Kumar, M.~S.~N., Sandell, G., et al.\ 2007, \mnras, 374, 29 



                                            
\bibitem[Dobashi et al.(2005)]{dob05} Dobashi, K., Uehara, H., Kandori, R., et al.\ 2005, \pasj, 57, 1 


\bibitem[Evans et al.(2003)]{evans03} Evans, N.~J., II, Allen, L.~E., Blake, G.~A., et al.\ 2003, \pasp, 115, 965 

\bibitem[Fich \& Blitz(1984)]{fic84} Fich, M., \& Blitz, L.\ 1984, \apj, 279, 125 

\bibitem[Flaherty et al.(2007)]{fla07} Flaherty, K.~M., Pipher, J.~L., Megeath, S.~T., et al.\ 2007, \apj, 663, 1069 

\bibitem[Frerking et al.(1982)]{frerking82} Frerking, M.~A., Langer, W.~D., \& Wilson, R.~W.\ 1982, \apj, 262, 590 

\bibitem[Goss \& Shaver(1970)]{gos70} Goss, W.~M., \& Shaver, P.~A.\ 1970, 
                                      Australian Journal of Physics Astrophysical Supplement, 14, 1 

                                      
\bibitem[Gutermuth et al.(2009)]{gut09} Gutermuth, R. A., Megeath, S. T., Myers, P. C., et al.\ 2009, \apjs, 184, 18 

\bibitem[Gutermuth et al.(2010)]{gut10} Gutermuth, R.~A., Megeath, S.~T., Myers, P.~C., et al.\ 2010, \apjs, 189, 352 

\bibitem[Hewett et al.(2006)]{hewett06} Hewett, P.~C., Warren, S.~J., Leggett, S.~K., \& Hodgkin, S.~T.\ 2006,
                                        \mnras, 367, 454 

\bibitem[Hillenbrand et al.(1992)]{hillenbrand92} Hillenbrand, L.~A., Strom, S.~E., Vrba, F.~J., \& Keene, J.\ 1992, 
                                                  \apj, 397, 613 
                                        
\bibitem[Hodgkin et al.(2009)]{hod09} Hodgkin, S.~T., Irwin, M.~J., Hewett, P.~C., \& Warren, S.~J.\ 2009, \mnras,
                                      394, 675 

\bibitem[Jenkins \& Savage(1974)]{jenkins74} Jenkins, E.~B., \& Savage, B.~D.\ 1974, \apj, 187, 243 
                                      
\bibitem[Kainulainen et al.(2007)]{kai07} Kainulainen, J., Lehtinen, K., V{\"a}is{\"a}nen, P., Bronfman, L.,
                                          \& Knude, J.\ 2007, \aap, 463, 1029 

\bibitem[K{\"o}nyves et al.(2010)]{kon10} K{\"o}nyves, V., Andr{\'e}, P., Men'shchikov, A., et al.\ 2010, \aap, 518, L106 

\bibitem[Krumholz et al.(2011)]{krumholz11} Krumholz, M.~R., Leroy, A.~K., \& McKee, C.~F.\ 2011, \apj, 731, 25 

\bibitem[Kuhn et al.(2010)]{kuh10} Kuhn, M.~A., Getman, K.~V., Feigelson, E.~D., et al.\ 2010, \apj, 725, 2485 

\bibitem[Kurtz et al.(1999)]{kurtz99} Kurtz, S.~E., Watson, A.~M., Hofner, P., \& Otte, B.\ 1999, \apj, 514, 232 

\bibitem[Lada(1987)]{lada87} Lada, C.~J.\ 1987, Star Forming Regions, 115, 1 

\bibitem[Lada \& Adams(1992)]{lada92} Lada, C.~J., \& Adams, F.~C.\ 1992, \apj, 393, 278 

\bibitem[Lada \& Lada(2003)]{lada03} Lada, C.~J., \& Lada, E.~A.\ 2003, \araa, 41, 57 

\bibitem[Lada et al.(1994)]{lad94} Lada, C.~J., Lada, E.~A., Clemens, D.~P., \& Bally, J.\ 1994, \apj, 429, 694 

\bibitem[Lilley(1955)]{lilley55} Lilley, A.~E.\ 1955, \apj, 121, 559 

\bibitem[Martins et al.(2005)]{martins05} Martins, F., Schaerer, D., \& Hillier, D.~J.\ 2005, \aap, 436, 1049 

\bibitem[Mathis(1990)]{mathis90} Mathis, J.~S.\ 1990, \araa, 28, 37 

\bibitem[Maury et al.(2011)]{mau11} Maury, A.~J., Andr{\'e}, P., Men'shchikov, A., K{\"o}nyves, V., \& Bontemps, S.\
                                    2011, \aap, 535, A77 

\bibitem[Meyer et al.(1997)]{meyer97} Meyer, M.~R., Calvet, N., \& Hillenbrand, L.~A.\ 1997, \aj, 114, 288 

\bibitem[Mezger \& Henderson(1967)]{mez67} Mezger, P.~G., \& Henderson, A.~P.\ 1967, \apj, 147, 471 

\bibitem[Mezger et al.(1967)]{mezEtAl67} Mezger, P.~G., Schraml, J., \& Terzian, Y.\ 1967, \apj, 150, 807 

\bibitem[Moran(1983)]{mor83} Moran, J.~M.\ 1983, Rev. Mexicana Astron. Astrofis., 7, 95
               
\bibitem[Myers(2009)]{mye09} Myers, P.~C.\ 2009, \apj, 700, 1609 

\bibitem[Nakamura et al.(2011)]{nak11} Nakamura, F., Sugitani, K., Shimajiri, Y., et al.\ 2011, \apj, 737, 56 

\bibitem[Ogura(2010)]{ogu10} Ogura, K.\ 2010, Astronomical Society of India Conference Series, 1, 19 

\bibitem[Ojha et al.(2004a)]{ojh04a} Ojha, D.~K., Tamura, M., Nakajima, Y., et al.\ 2004, \apj, 608, 797 

\bibitem[Ojha et al.(2004b)]{ojh04b} Ojha, D.~K., Tamura, M., Nakajima, Y., et al.\ 2004, \apj, 616, 1042 

\bibitem[Panagia(1973)]{pan73} Panagia, N.\ 1973, \aj, 78, 929 

\bibitem[Parker et al.(2005)]{parker05} Parker, Q.~A., Phillipps, S., Pierce, M.~J., et al.\ 2005, \mnras, 362, 689 

\bibitem[Pilbratt et al.(2010)]{pil10} Pilbratt, G.~L., Riedinger, J.~R., Passvogel, T., et al.\ 2010, \aap, 518, L1 

\bibitem[Pirogov et al.(2012)]{pir12} Pirogov, L., Zinchenko, I., Ojha, D.~K., \& Ghosh, S.~K.\ 2012,
                                      The Astronomer's Telegram, 4236, 1 

\bibitem[Pirogov et al.(2013)]{pirogov13} Pirogov, L., Ojha, D.~K., Thomasson, M., et al.\ 2013, \mnras (accepted); 
                                          arXiv:1309.6188 
                                      
\bibitem[Radhakrishnan et al.(1972)]{radhakrishnan72} Radhakrishnan, V., Goss, W.~M., Murray, J.~D., \& Brooks, J.~W.\ 
                                                      1972, \apjs, 24, 49

\bibitem[Rieke \& Lebofsky(1985)]{rieke85} Rieke, G.~H., \& Lebofsky, M.~J.\ 1985, \apj, 288, 618 


                                          

\bibitem[Rodney \& Reipurth(2008)]{rod08} Rodney, S.~A., \& Reipurth, B.\ 2008, Handbook of Star Forming Regions,
                                          Volume II, 683 

\bibitem[Rodr{\'{\i}}guez et al.(2010)]{rod10} Rodr{\'{\i}}guez, L.~F., Rodney, S.~A., \& Reipurth, B.\ 2010, \aj, 140, 968 

\bibitem[Schmeja et al.(2008)]{schmeja08} Schmeja, S., Kumar, M.~S.~N., \& Ferreira, B.\ 2008, \mnras, 389, 1209 

\bibitem[Schmeja(2011)]{schmeja11} Schmeja, S.\ 2011, Astronomische Nachrichten, 332, 172 

\bibitem[Schneider et al.(2012)]{sch12} Schneider, N., Csengeri, T., Hennemann, M., et al.\ 2012, \aap, 540, L11 

\bibitem[Shaver \& Goss(1970)]{sha70} Shaver, P.~A., \& Goss, W.~M.\ 1970,
                                      Australian Journal of Physics Astrophysical Supplement, 14, 133 

\bibitem[Shuping et al.(1999)]{shu99} Shuping, R.~Y., Snow, T.~P., Crutcher, R., \& Lutz, B.~L.\ 1999, \apj, 520, 149 
                                      
\bibitem[Shuping et al.(2012)]{shu12} Shuping, R.~Y., Vacca, W.~D., Kassis, M., \& Yu, K.~C.\ 2012, \aj, 144, 116 

\bibitem[Smith et al.(1985)]{smi85} Smith, J., Bentley, A., Castelaz, M., et al.\ 1985, \apj, 291, 571 

\bibitem[Spitzer(1978)]{spi78} Spitzer, L. 1978, Physical Processes in the Interstellar Medium 
                               (New York: Wiley-Interscience) 

\bibitem[Stahler \& Palla(2005)]{sta05} Stahler, S.~W., \& Palla, F.\ 2005,
                                        The Formation of Stars,
                                        by Steven W.~Stahler, Francesco Palla,
                                        pp.~865.~ISBN 3-527-40559-3.~Wiley-VCH , January 2005.  

\bibitem[Str{\"o}mgren(1939)]{str39} Str{\"o}mgren, B.\ 1939, \apj, 89, 526 

\bibitem[Swarup et al.(1991)]{swa91} Swarup, G., Ananthakrishnan, S., Kapahi, V.~K., et al.\ 1991,
                                     Current Science, Vol.~60, NO.2/JAN25, P.~95, 1991, 60, 95 

\bibitem[Tenorio-Tagle(1979)]{teno79} Tenorio-Tagle, G.\ 1979, \aap, 71, 59 
                                     
\bibitem[Vallee(1987)]{val87} Vallee, J.~P.\ 1987, \aap, 178, 237 

\bibitem[Vallee et al.(1992)]{val92} Vallee, J.~P., Guilloteau, S., \& MacLeod, J.~M.\ 1992, \aap, 266, 520 

\bibitem[Westerhout(1958)]{wes58} Westerhout, G.\ 1958, \bain, 14, 215 

\bibitem[Ward-Thompson \& Whitworth(2011)]{war11} Ward-Thompson, D., \& Whitworth, A.~P.\ 2011,
                                                  An Introduction to Star Formation
                                                  by Derek Ward-Thompson and Anthony P.~Whitworth.
                                                  ~Cambridge University Press, 2011.~ISBN: 9780521630306   

\bibitem[Waskett et al.(2007)]{was07} Waskett, T.~J., Sibthorpe, B., Griffin, M.~J., \& Chanial, P.~F.\ 2007,
                                      \mnras, 381, 1583 



\bibitem[Whitworth(1979)]{whit79} Whitworth, A.\ 1979, \mnras, 186, 59 

\bibitem[Quireza et al.(2006)]{qui06} Quireza, C., Rood, R.~T., Bania, T.~M., Balser, D.~S., \& 
                                      Maciel, W.~J.\ 2006, \apj, 653, 1226 

\bibitem[Yasui et al.(2008)]{yasui08} Yasui, C., Kobayashi, N., Tokunaga, A.~T., Terada, H., \& Saito, M.\ 2008, 
                                      \apj, 675, 443 

\bibitem[Zeilik \& Lada(1978)]{zei78} Zeilik, M., II, \& Lada, C.~J.\ 1978, \apj, 222, 896 

\bibitem[Zhu et al.(2006)]{zhu06} Zhu, L., Wu, Y.-F., \& Wei, Y.\ 2006, Chinese J. Astron. Astrophys., 6, 61 


\end{thebibliography}
\end{document}